\documentclass[sigconf]{acmart}
\AtBeginDocument{%
  }

\setcopyright{acmlicensed}
\copyrightyear{2025}
\acmYear{2025}
\acmDOI{XXXXXXX.XXXXXXX}
\acmConference[KDD '25] {Proceedings of the ACM SIGKDD Conference on Knowledge Discovery and Data Mining 2025} {August 3-7, 2025}{Toronto, Canada.}
\acmBooktitle{Proceedings of the ACM SIGKDD Conference on Knowledge Discovery and Data Mining 2025 (KDD '25), August 3-7, 2025, Toronto, Canada}
\acmISBN{978-1-4503-XXXX-X/2018/06}

\usepackage{times}
\usepackage{soul}
\usepackage{url}
\usepackage[utf8]{inputenc}
\usepackage{graphicx}
\usepackage{amsmath}
\usepackage{amsthm}
\usepackage{booktabs}
\usepackage[switch]{lineno}
\usepackage[ruled, vlined, boxed, linesnumbered]{algorithm2e}
\usepackage{multirow}

\begin{document}

%%
%% The "title" command has an optional parameter,
%% allowing the author to define a "short title" to be used in page headers.
%%%% Subgraph Attacks Against Graph-based Fake News Detectors Using Structural Information
\title{Robustness Evaluation of Graph-based News Detection Using Network Structural Information}

%%
%% The "author" command and its associated commands are used to define
%% the authors and their affiliations.
%% Of note is the shared affiliation of the first two authors, and the
%% "authornote" and "authornotemark" commands
%% used to denote shared contribution to the research.

\author{Xianghua Zeng}
\email{zengxianghua@buaa.edu.cn}
\affiliation{
    \institution{State Key Laboratory of Software Development Environment, Beihang University}
    \city{Beijing}
    \country{China}
}

\author{Hao Peng}
\authornote{Corresponding authors.}
\email{penghao@buaa.edu.cn}
\affiliation{
    \institution{State Key Laboratory of Software Development Environment, Beihang University}
    \city{Beijing}
    \country{China}
}

\author{Angsheng Li}
\authornotemark[1]
\email{angsheng@buaa.edu.cn}
\affiliation{
    \institution{State Key Laboratory of Software Development Environment, Beihang University}
    \city{Beijing}
    \country{China}
}

\begin{abstract}
Although Graph Neural Networks (GNNs) have shown promising potential in fake news detection, they remain highly vulnerable to adversarial manipulations within social networks.
Existing methods primarily establish connections between malicious accounts and individual target news to investigate the vulnerability of graph-based detectors, while they neglect the structural relationships surrounding targets, limiting their effectiveness in robustness evaluation.
In this work, we propose a novel \textbf{S}tructural \textbf{I}nformation principles-guided \textbf{A}dversarial \textbf{A}ttack \textbf{F}ramework, namely \textbf{SI2AF}, which effectively challenges graph-based detectors and further probes their detection robustness.
Specifically, structural entropy is introduced to quantify the dynamic uncertainty in social engagements and identify hierarchical communities that encompass all user accounts and news posts.
An influence metric is presented to measure each account's probability of engaging in random interactions, facilitating the design of multiple agents that manage distinct malicious accounts.
For each target news, three attack strategies are developed through multi-agent collaboration within the associated subgraph to optimize evasion against black-box detectors.
By incorporating the adversarial manipulations generated by SI2AF, we enrich the original network structure and refine graph-based detectors to improve their robustness against adversarial attacks. 
Extensive evaluations demonstrate that SI2AF significantly outperforms state-of-the-art baselines in attack effectiveness with an average improvement of $16.71\%$, and enhances GNN-based detection robustness by $41.54\%$ on average.
\end{abstract}

%%
%% The code below is generated by the tool at http://dl.acm.org/ccs.cfm.
%% Please copy and paste the code instead of the example below.
%%
% \begin{CCSXML}
% <ccs2012>
%  <concept>
%   <concept_id>00000000.0000000.0000000</concept_id>
%   <concept_desc>Do Not Use This Code, Generate the Correct Terms for Your Paper</concept_desc>
%   <concept_significance>500</concept_significance>
%  </concept>
%  <concept>
%   <concept_id>00000000.00000000.00000000</concept_id>
%   <concept_desc>Do Not Use This Code, Generate the Correct Terms for Your Paper</concept_desc>
%   <concept_significance>300</concept_significance>
%  </concept>
%  <concept>
%   <concept_id>00000000.00000000.00000000</concept_id>
%   <concept_desc>Do Not Use This Code, Generate the Correct Terms for Your Paper</concept_desc>
%   <concept_significance>100</concept_significance>
%  </concept>
%  <concept>
%   <concept_id>00000000.00000000.00000000</concept_id>
%   <concept_desc>Do Not Use This Code, Generate the Correct Terms for Your Paper</concept_desc>
%   <concept_significance>100</concept_significance>
%  </concept>
% </ccs2012>
% \end{CCSXML}

% \ccsdesc[500]{Do Not Use This Code~Generate the Correct Terms for Your Paper}
% \ccsdesc[300]{Do Not Use This Code~Generate the Correct Terms for Your Paper}
% \ccsdesc{Do Not Use This Code~Generate the Correct Terms for Your Paper}
% \ccsdesc[100]{Do Not Use This Code~Generate the Correct Terms for Your Paper}

%%
%% Keywords. The author(s) should pick words that accurately describe
%% the work being presented. Separate the keywords with commas.
%% A "teaser" image appears between the author and affiliation
%% information and the body of the document, and typically spans the
%% page.
% \begin{teaserfigure}
%   \includegraphics[width=\textwidth]{sampleteaser}
%   \caption{Seattle Mariners at Spring Training, 2010.}
%   \Description{Enjoying the baseball game from the third-base
%   seats. Ichiro Suzuki preparing to bat.}
%   \label{fig:teaser}
% \end{teaserfigure}

% \received{20 February 2007}
% \received[revised]{12 March 2009}
% \received[accepted]{5 June 2009}

%%
%% This command processes the author and affiliation and title
%% information and builds the first part of the formatted document.
\maketitle

\section{Introduction}\label{Introduction}
Recent studies \cite{shu2017fake, chen2022combating} have highlighted that the widespread growth of social media has accelerated the dissemination of misinformation and fake news. 
This phenomenon not only undermines public trust but also has detrimental effects on critical domains such as politics \cite{aral2019protecting}, economics \cite{di2021fake}, and public safety \cite{vosoughi2018spread}. 
Unlike traditional news articles, fake content on social platforms poses unique challenges due to its intentionally misleading nature, rapid spread, and high costs associated with expert verification \cite{ren2020adversarial}, which necessitates the development of automated detection mechanisms.
Traditional machine learning detectors \cite{ruchansky2017csi,shu2019defend} employing natural language processing techniques aim to identify fake content to curb the spread of misinformation online.
However, these approaches face efficiency limitations, especially in their capacity to account for the unique dispersion structures and complex spreading behaviors of misinformation \cite{bian2020rumor}.
To bridge this gap, Graph Neural Network (GNN)-based detectors \cite{lu2020gcan,nguyen2020fang} have emerged, providing a more precise analysis of the intricate structural patterns in rumor dissemination, thereby notably enhancing detection accuracy.

Despite advancements in GNN techniques, current graph-based detectors remain susceptible to adversarial manipulations \cite{dai2018adversarial}. 
While recent work has extensively explored the resilience of NLP-based detectors \cite{horne2019robust,he2021petgen}, the robustness of graph-based detectors remains largely under-researched.
The Malcom framework was developed to systematically probe and exploit vulnerabilities in advanced fake news detection systems through the generation of adversarial comments \cite{le2020malcom}.
Additionally, a reinforcement learning-based attack strategy was designed to identify specific weaknesses in sophisticated graph-based rumor detectors \cite{lyu2023interpretable}.
The gradient-based GAFSI framework \cite{zhu2024general} was introduced to enable general adversarial attacks against black-box detectors across various graph structures.
Recognizing important social interactions and diverse fraudster types, a multi-agent reinforcement learning adversarial attack framework was proposed, employing three types of fraudsters to investigate the vulnerabilities of graph-based detectors in adversarial scenarios thoroughly \cite{wang2023attacking}.
Nevertheless, these methods primarily focus on associating malicious accounts with individual target news, overlooking the underlying structural relationships within the social network, which play a pivotal role in the propagation of misinformation.

\begin{figure}[t]
   \centering
    \includegraphics[width=0.77\columnwidth]{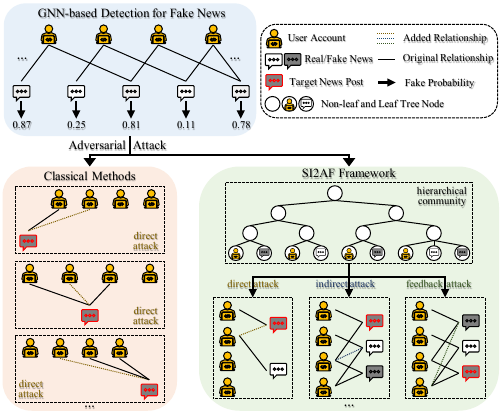}
    \vspace{-0.4cm}
    \caption{Comparative illustration between classical methods and our framework. SI2AF minimizes dynamic uncertainty in social engagements and identifies an associated subgraph to strategically establish connections with both target and non-target posts, resulting in significantly enhanced attack effectiveness.}
    \vspace{-0.9cm}
    \label{attack} 
\end{figure}

This work integrates the network's structural information into SI2AF, a comprehensive attack framework designed to enhance the understanding of misinformation dynamics and assess the robustness of graph-based detectors.
Initially, we extract user accounts and news posts from historical engagements to construct a bipartite user-post graph, where interactions between users and posts are modeled as random walks among graph vertices, with their dynamic uncertainties quantified through structural entropy. 
Subsequently, we minimize the high-dimensional structural entropy of this graph to identify a hierarchical community structure for all vertices, referred to as the optimal encoding tree.
Each tree node corresponds to a vertex community, and an associated subgraph captures the frequent engagements between user and post vertices within this community.
Furthermore, we present an influence metric derived from structural entropy to measure the likelihood of each account’s participation in random engagements within the bipartite graph, reflecting its potential impact on the information flow.
Building on this metric, we categorize user accounts into genuine accounts and distinct malicious groups with varying levels of influence, each managed by a separate decision-making agent.
For each target post, we develop three attack strategies through multi-agent collaboration within the associated user-post subgraph, aiming to maximize its evasion against black-box detectors.
Figure \ref{attack} illustrates a comparative analysis of attacks against a graph-based fake news detector using traditional methods and our SI2AF framework.
By incorporating the generated manipulations, we enrich the structural relationships within the user-post graph and further refine graph-based models to improve detection robustness against adversarial attacks.
Comprehensive experiments conducted on two real-world datasets demonstrate that SI2AF significantly outperforms state-of-the-art baselines regarding attack effectiveness and effectively enhances the robustness of graph-based detection. 
% The source codes and detailed results are publicly available via an anonymous link\footnote{\url{https://anonymous.4open.science/r/b485a0b/}} for verification and comparison.
Our contributions can be summarized as follows:

$\bullet$ A novel adversarial attack framework that leverages network structural information is proposed to effectively obfuscate graph-based news detectors and evaluate their robustness.

$\bullet$ An influence metric is presented to quantify the likelihood of user accounts' participation in random engagements, enabling the design of multiple agents that manage distinct malicious accounts.

$\bullet$ Three attack strategies based on multi-agent coordination within bipartite user-post subgraphs are developed to optimize evasion against graph-based detectors.

$\bullet$ Comparative evaluations demonstrate that SI2AF significantly improves attack effectiveness by $16.71\%$ and enhances graph-based detection robustness by $41.54\%$ at average.

\section{Related Work}\label{Related Work}

\subsection{Adversarial Attack on Graph Neural Networks}
Various adversarial techniques have been developed to attack graph-based detectors via edge perturbation, aiming to probe their robustness.
Nettack \cite{zugner2018adversarial} introduced the first targeted attack utilizing incremental computation and greedy-based edge perturbations, optimizing the attack strategy step by step.
SGA \cite{li2021adversarial} improved attack efficiency on large-scale graphs by incorporating a subgraph construction process to misclassify targeted nodes.
The minimum-budget topology attack strategy \cite{zhang2023minimum} was designed to determine the smallest amount of perturbation necessary to compromise each node successfully.
EA-PGD \cite{shang2023transferable} introduced transferable adversarial attacks to perform edge perturbations on heterogeneous graph structures.
Despite their potential successes, these methods disrupt the propagation structure, resulting in insufficiency when attacking social news detectors.
Recent advances \cite{wang2023attacking} have led to developing a multi-agent coordination framework on three types of malicious accounts to disrupt GNN-based fake news detection.
The gradient-based GAFSI method \cite{zhu2024general} has successfully executed general adversarial attacks against detectors across various graph structures.
However, these approaches typically use malicious accounts to engage with individual target posts, neglecting to account for the structural relationships between posts, which limits their attack strategies and reduces the effectiveness of the attack. 
In contrast, our work novelly integrates structural information principles to design a range of subgraph-based attack strategies that are both more nuanced and effective than previous methods. 
The attack strategies outlined above are summarized in Table \ref{table: attack summary}, which evaluates whether they target social news detection, model distinct groups of malicious accounts, and consider the structural relationships among target posts.

\vspace{-0.2cm}
\begin{table}[h]
\centering
\caption{Summary of attacks against GNN-based detectors.}
\vspace{-0.4cm}
\label{table: attack summary}
\resizebox{\columnwidth}{!}{
\begin{tabular}{c|ccc}
\hline
\textbf{Attack Method}  & \textbf{News Detection} & \textbf{Distinct Group} & \textbf{Structural Relationship} \\ \hline
Nettack \cite{zugner2018adversarial} & $\times$ & $\times$ & $\times$ \\
SGA \cite{li2021adversarial} & $\times$ & $\times$ & $\checkmark$ \\
MiBTack \cite{zhang2023minimum} & $\times$ & $\times$ & $\times$ \\
EA-PGD \cite{shang2023transferable} & $\times$ & $\times$ & $\times$ \\ \hline
MARL \cite{wang2023attacking} & $\checkmark$ & $\checkmark$ & $\times$ \\
GAFSI \cite{zhu2024general} & $\checkmark$ & $\times$ & $\times$ \\ \hline
SIASF (Ours) & $\checkmark$ & $\checkmark$ & $\checkmark$ \\ \hline
\end{tabular}}
\end{table}
\vspace{-0.5cm}

\subsection{Structural Information Principles}
In 2016, a significant advancement was achieved with the introduction of structural information principles, particularly structural entropy and partitioning trees, as proposed in \cite{li2016structural}. 
These concepts facilitated the measurement of network complexity and laid the groundwork for identifying hierarchical communities within complex systems. 
Building on these principles, researchers minimize high-dimensional structure entropy to classify cancer cell subtypes \cite{li2016three} and decode topologically associating domains within Hi-C data \cite{li2018decoding}.
Subsequently, structural information guided graph learning methods demonstrate promising results in node and graph classification tasks \cite{wu2022structural,zou2023se}.
Further advancements apply adaptive partitioning of social accounts based on encoding trees to model adversarial behaviors and enable unsupervised detection of social bots \cite{peng2024unsupervised,zeng2024adversarial}. 
More recent studies \cite{zeng2023effective,zeng2023hierarchical,zeng2025effective} have focused on developing efficient decision-making algorithms within partitioning trees that structure the state or action spaces.

\section{Preliminaries}\label{Problem Formulation}
In this section, we begin with a definition of GNN-based fake news detection, followed by a description of adversarial attacks targeting these detectors, and conclude with an introduction to the structural information principles.

\subsection{GNN-based Fake News Detection}
A user-post graph, denoted as $G_{up}$, is a bipartite graph defined by the tuple $\{U,P,E_{up},X_u,X_p,Y\}$, where $U=(u_0,\dots,u_m)$ represents the set of users and $P=(p_0,\dots,p_n)$ denotes the set of news posts.
The feature matrices for users and posts are represented by $X_u$ and $X_p$, respectively. 
The label set $Y$ contains post labels, where $1$ signifies fake news and $0$ indicates real news.
An edge $e_{ij}=(u_i,p_j) \in E_{up}$ implies social engagement between user $u_i$ and post $p_j$.

Within a standard detection framework, a graph neural network (GNN) $f_\theta$ processes $G_{up}$ by recursively aggregating information from neighboring vertices to obtain a representation $h_p$ for each post $p$.
To classify a given news post $p \in P$, the GNN representation $h_p$ is input into a classifier $f$, which maps $h_p$ to a predicted label $\hat{y} \in (0,1)$.
The cross-entropy loss for $P$ is formulated as follows:
\begin{equation} \label{equ: ce loss}
    \mathcal{L}_{GNN}\left(G_{up},f_\theta\right) = \sum_{p_i \in P} \left[- \log \left(y_i \cdot \sigma\left(f_\theta\left(X_p,E_{up}\right)_{p_i}\right)\right)\right]\text{,}
\end{equation}
where $\sigma$ denotes the sigmoid function used for the binary classification of news posts.

\subsection{Attacks against GNN-based News Detectors}
An adversarial attack against GNN-based detectors aims to alter the classification outcomes of target news posts $P_t \subset P$ by leveraging malicious accounts $U_m \subset U$ that disseminate new posts. 
In this study, the GNN model $f_\theta$ is initially trained on a clean dataset, and it is assumed that the model's parameters remain unknown during the attack process.
The attack objective is to maximize the misclassification rate among $P_t$, which is expressed as follows:
\begin{equation}
    \begin{aligned}
        & \max_{E_{up}^\prime} \sum_{p_i \in P_t} \mathbf{1}\left(f_{\theta^*}\left(X_p,E_{up}^\prime\right)_{p_i} \neq y_i\right)\text{,} \\
        \text{s.t.} \quad &\theta^* = \arg \min \mathcal{L}_{GNN}\left(G_{up},f_\theta\right)\text{,}\quad 
        \left|U_m\right| \leq \Delta_u \text{,}
    \end{aligned}
\end{equation}
where $E_{up}^\prime$ represents the manipulated edges, and $\Delta_u$ denotes the attack budget, defined as the maximum number of controlled users.

\subsection{Structural Information Principles}
In an undirected graph $G=(V, E)$, a disjoint partition of all vertices is denoted as ${V_0, V_1, \dots}$, with each subset $V_i$ representing as a vertex community. 
These primary communities can be further subdivided into smaller sub-communities, forming a hierarchical community structure. 
The concept of structural entropy \cite{li2016structural} quantifies the dynamic uncertainty encountered during a random walk between vertices within this hierarchical structure. 

In the absence of a hierarchical community structure, the one-dimensional structural entropy $H^1(G)$ of the graph $G$ is analogous to Shannon entropy \cite{shannon1948mathematical} and is calculated based on the distribution of vertex degrees $d_v$ as follows:
\begin{equation}
    H^1(G) = - \sum_{v \in V} d_v \cdot \log d_v\text{.}
\end{equation}

In this work, we define the hierarchical community structure used in SI2AF as an encoding tree with the following properties:
1) The root node $\lambda$ corresponds to the entire vertex set $V$, such that $V_\lambda=V$.
2) Each leaf node $\nu$ corresponds to an individual vertex $v \in V$, with $V_\nu=\{v\}$.
3) Each intermediate node $\alpha$ (neither root nor leaf) corresponds to a subset of vertices $V_\alpha$, and its parent node is marked as $\alpha^-$.
4) For each non-leaf node $\alpha$, the number of its child nodes is assumed as $l_\alpha$, and its $i$-th child is specified as $\alpha^{\langle i \rangle}$. 
The subsets $V_{\alpha^{\langle i \rangle}}$ are mutually exclusive and collectively exhaustive, satisfying $V_\alpha=\bigcup_{i=1}^{l_\alpha}V_{ \alpha^{\langle i \rangle}}$ and $V_{\alpha^{\langle i \rangle}} \cap V_{\alpha^{\langle j \rangle}} = \emptyset$ for any $i \neq j$.

The encoding tree $ T$ significantly reduces the dynamical uncertainty within the graph $G$, and the high-dimensional structural entropy quantifies the residual uncertainty.
The entropy assigned to a non-root node $\alpha$ represents the uncertainty associated with a random walk transitioning from the parent community $V_{\alpha^{-}}$ to the child community $V_\alpha$, as detailed as follows:
\begin{equation}\label{kd_se_node}
    H^{T}(G;\alpha)=-\frac{g_{\alpha}}{\mathcal{V}_\lambda} \log _{2} \frac{\mathcal{V}_{\alpha}}{\mathcal{V}_{\alpha^{-}}}\text{,}
\end{equation}
where $\mathcal{V}_\alpha$ is the volume of $V_\alpha$, $\mathcal{V}_\alpha = \sum_{v \in V_\alpha}d_v$.
The item $g_\alpha$ denotes the cumulative weight of all edges connecting vertices within $V_\alpha$ to vertices outside $V_\alpha$.
The $K$-dimensional structural entropy is defined as follows:
\begin{equation}\label{kd_se}
     H^T(G) = \sum_{\alpha \in T, \alpha \neq \lambda}H^{T}(G;\alpha)\text{,} \quad H^{K}(G) = \min_{T}\left\{H^T(G)\right\}\text{,}
\end{equation}
where $T$ ranges over all encoding trees with heights at most $K > 1$.

\section{Methodology}\label{State Abstraction}
\begin{figure*}[t]
   \centering
    \includegraphics[width=\textwidth]{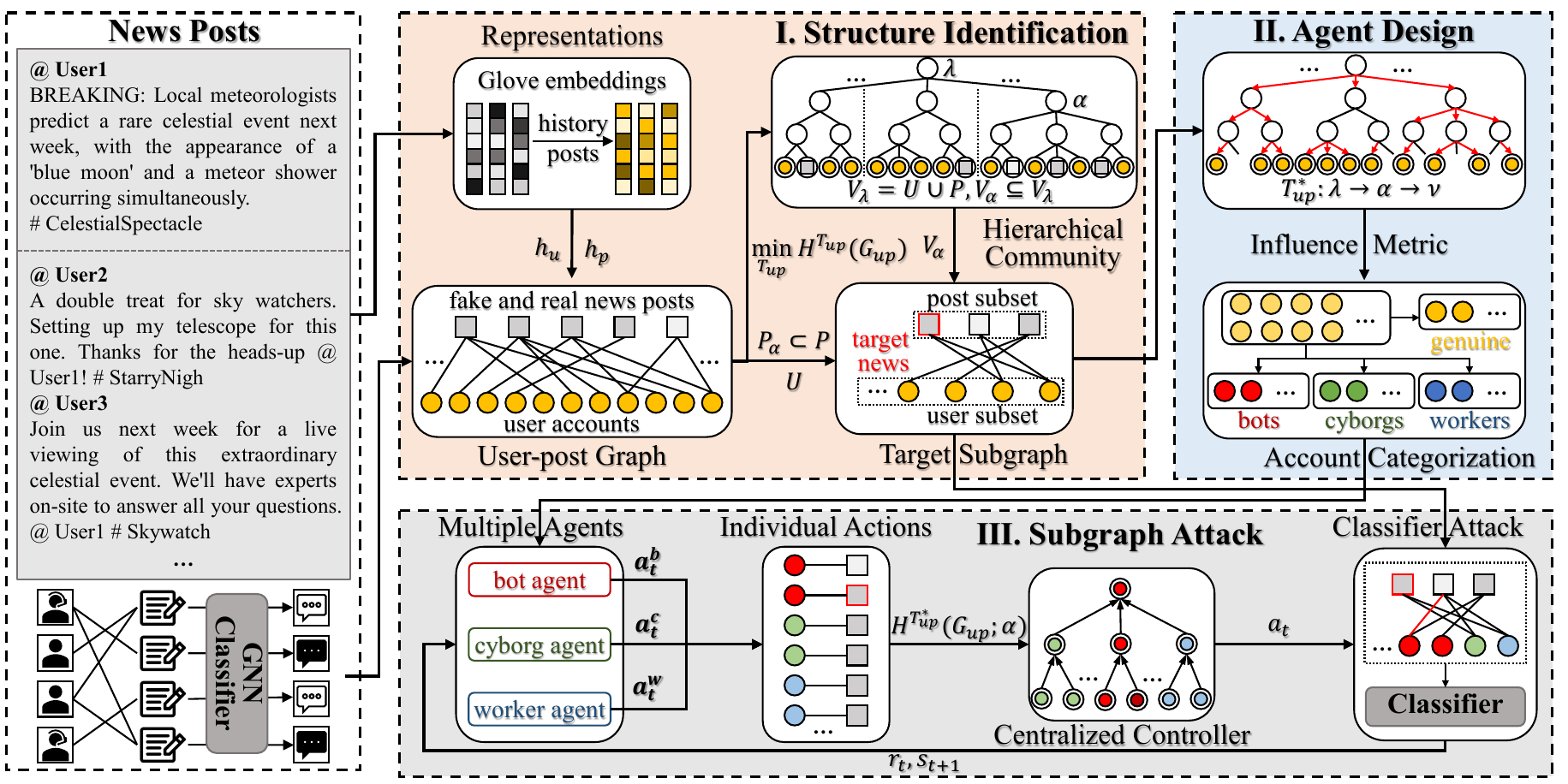}
    \vspace{-0.65cm}
    \caption{Detailed design of our proposed SI2AF framework.}
    \vspace{-0.45cm}
    \label{fig: SI2AF framework} 
\end{figure*}

In this work, we leverage structural information in social networks to identify the hierarchical community structure encompassing user accounts and news posts and further utilize multi-agent coordination to achieve effective attacks against GNN-based news detectors.
As illustrated in Figure \ref{fig: SI2AF framework}, our SI2AF framework consists of three primary modules: hierarchical structure identification, multiple agent design, and target subgraph attack.
During the structure identification module, we construct a bipartite user-post graph from historical engagements and generate its optimal encoding tree, representing the hierarchical community structure of all users and posts.
In the agent design module, we present an influence metric using structural entropy to evaluate user accounts, categorizing them into distinct types of malicious and genuine accounts.
We coordinate multiple agents for each target post to establish new connections with both target and non-target posts within the associated subgraph, aiming to optimize evasion under GNN-based detection models.

\subsection{Hierarchical Structure Identification} \label{hierarchical structure extraction}
In contrast to previous studies \cite{wang2023attacking, zhu2024general}, which independently analyze individual target news, we minimize the dynamic uncertainty in social engagements to identify a hierarchical community structure of social accounts and news posts, thereby facilitating effective subgraph attacks within the SI2AF framework.

To this end, we begin by extracting historical engagements between user accounts $U$ and news posts $P$ to construct an undirected bipartite user-post graph $G_{up}=(U,P,E_{up})$.
Following the methodology described by \cite{dou2021user}, we employ a pre-trained language model \cite{pennington2014glove} to obtain user representations $X_u$ by embedding their historical posts.
Similarly, we create post representations $X_p$ by embedding the content of each news post.

For each edge $e_{ij} = (u_i,p_j) \in E_{up}$, we calculate the cosine similarity between the representations $h_{u_i} \in X_u$ and $h_{p_j} \in X_p$, a standard measure for capturing semantic similarity in embedding spaces.
The resulting similarity score is used to compute the edge weight $w_{ij} \in [0, 1]$ as follows:
\begin{equation}
    w_{ij} = \frac{1}{2} \left(\operatorname{cos}\left(h_{u_i},h_{p_j}\right) + 1\right) \text{.}
\end{equation}
Intuitively, a higher weight $w_{ij}$ indicates greater relevance between user $u_i$ and post $p_j$, whereas a lower weight reflects dissimilarity.

In the bipartite graph $G_{up}$, we then model social engagements as random walks between user and post vertices, using structural entropy to quantify the dynamic uncertainty of these interactions.
This entropy quantifies the minimum amount of information (in bits) required to determine accessible users or posts during a random social engagement.
By minimizing the high-dimensional entropy of $G_{up}$, we generate its optimal encoding tree, which captures the hierarchical community structure of user accounts $U$ and news posts $P$.
We start by initializing a single-layer encoding tree $T_{up}$ for $G_{up}$, where each leaf node $\nu$ has the tree root $\lambda$ as its parent, denoted as $\nu^-=\lambda$.
Using the HCSE algorithm \cite{pan2021information}, we apply two operators, \textit{stretch} and \textit{compress}, to iteratively and greedily optimize the encoding tree $T_{up}$ from a single layer to $K$ layers, ultimately yielding the optimal $K$-layer encoding tree $T^*_{up}$.
In the tree $T^*_{up}$, the root node $\lambda$ corresponds to the union of user and post sets, $V_\lambda = U \cup P$.
Each leaf node $\nu$ corresponds to a singleton containing an individual user or post, while intermediate nodes correspond to communities at various hierarchical levels.

Finally, for each target post $p \in P_t$, we extract its corresponding $k$-layer community $V_\alpha$, compressing the user subset $U_\alpha \subset U$ and post subset $P_\alpha \subset P$ at the $k$-th hierarchical level in $T^*_{up}$.
We extend the user subset $U_\alpha$ to include the entire set $U$, and derive the associated bipartite subgraph $G_\alpha=(U,P_\alpha,E_\alpha^{up})$.
The extended vertex subset consists of two components: the entire account set $U$ and the post subset $P_\alpha \subset P$.
The edge subset $E_\alpha^{up}$ captures the local structural relationships between the accounts in $U$ and posts in $P_\alpha$, highlighting their interactions within the subgraph.

In this work, we set the height parameter $k$ as $K-1$ by default, enabling us to derive all targeted subgraphs from the vertex communities corresponding to the immediate children of the root node.

\subsection{Multiple Agent Design}
Building on this hierarchical community structure, we present a metric to measure each user account's network influence and design multiple cooperative agents to manage malicious accounts, taking into account their distinct influences and budgets.

In the encoding tree $T^*_{up}$, the structural entropy assigned to each non-root node $\alpha$ in Equation \ref{kd_se_node} measures the uncertainty of a random walk transitioning from the parent community $V_{\alpha^-}$ to its child community $V_\alpha$.
For any user $u \in U$, the probability of a random engagement reaching this user is determined by the cumulative entropy of all nodes $\alpha$ encountered along the path from the root node $\lambda$ to the leaf node $\nu$, where $V_\nu=\{u\}$.
Consequently, we define the influence metric $\mathcal{I}$ as a measure of each user account's likelihood of engaging in random interactions within $G_{up}$ as detailed below:
\begin{equation}\label{equ: influence metric}
    \mathcal{I}(G_{up};u) = \sum_{V_{\nu} \subseteq V_\alpha \subset V}{\left[-\frac{g_{\alpha}}{\mathcal{V}_\lambda} \log _{2} \frac{c \cdot \mathcal{V}_{\alpha}}{\mathcal{V}_{\alpha^{-}}}\right]}\text{,}
\end{equation}
where $c$ serves as an adjusting parameter that modulates the distribution of influence across all user accounts.

Prior research \cite{wang2023attacking} has categorized distinct malicious groups according to the number of news shares per account, which indicates their network influence. 
However, due to the sparsity of social networks, where most users are linked to only a single news post, this leads to a heavy-tailed sharing distribution, which can cause imbalances in influence-based categorizations.
By integrating content relevance and hierarchical community structure into our metric, we achieve a more nuanced differentiation between users who share the same number of posts, enhancing the precision of influence measurement compared to previous methods.
The following theorem demonstrates that, even in the context of an unweighted graph and a single-layer network structure, adjusting the parameter $c$ within the influence metric can reduce the occurrence of accounts with identical influence values, thus fostering a more balanced distribution of user influence.
In this particular scenario, the influence metric $\mathcal{I}(G_{up};u)$ for each user $u \in U$ depends exclusively on the user's vertex degree, which represents the number of connections the user has to different pieces of content. 
This influence metric is formally defined as follows:
\begin{equation}
    \mathcal{I}(G_{up};u) = -\frac{g_{\nu}}{\mathcal{V}_\lambda} \log _{2} \frac{c \cdot \mathcal{V}_\nu}{\mathcal{V}_\lambda} = -\frac{d_u}{\mathcal{V}_\lambda} \log _{2} \frac{c \cdot d_u}{\mathcal{V}_\lambda}\text{.}
\end{equation}
\begin{theorem} \label{theorem: 4.1}
    Let $x \in [1,\frac{b}{2}]$ be a positive random variable with a probability density function $q_0(x)$.
    Given the transformation $x^\prime = -\frac{x}{b} \cdot \left(\log_{2} \frac{c}{b} x\right)$, under the condition $0 < c \leq \frac{2}{e}$, the variable $x^\prime$ increases monotonically with the variable $x$, and its probability density function $q_1(x^\prime)$ satisfies:
    \begin{equation} \label{equ: y_pdf}
        0 \leq q_1(x^\prime) \leq \frac{b}{1 - \log_2 ec} \text{.}
    \end{equation}
\end{theorem}
\noindent A detailed proof of this theorem is provided in Appendix \ref{appendix: 4.1}.
The parameter $b$ denotes the sum of the total number of posts shared by all users and the total number of times these posts have been shared.
As a result, each user's individual sharing count $x$ is bounded by the range of $1 \leq x \leq \frac{b}{2}$.

% \begin{table}[t]
% \centering
% \caption{Property comparison of different malicious accounts.}
% \vspace{-0.3cm}
% \label{table: agent comparison}
% \resizebox{\columnwidth}{!}{
% \begin{tabular}{c|cc}
% \hline
% \textbf{Malicious Account} & \textbf{Network Influence} & \textbf{Budget Constraint} \\ \hline
% bots & low & high \\
% cyborgs & medium & medium \\
% crowded workers & high & low \\ \hline
% \end{tabular}}
% \vspace{-0.5cm}
% \end{table}

\begin{algorithm}[t]
\SetAlgoVlined
\KwIn{user account set $U$, bot budget $\Delta_b$, cyborg budget $\Delta_c$, worker budget $\Delta_w$}
\KwOut{bot set $U_b$, cyborg set $U_c$, worker set $U_w$}
% Sort users based on network influence metric
$U^\prime \gets$ sort $U$ by network influence using Equation \ref{equ: influence metric} \\
% Total budget calculation
$\Delta \gets \Delta_b + \Delta_c + \Delta_w$ \\
% Calculate partition sizes for each category
% $m_l \gets \lfloor \frac{\Delta_b}{\Delta} \rfloor$, $m_m \gets \lfloor \frac{\Delta_c}{\Delta} \rfloor$, $m_h \gets \lfloor \frac{\Delta_w}{\Delta} \rfloor$ \\
% Partition the user set into three categories
$U_l \gets U^\prime[: \lfloor \frac{\Delta_b}{\Delta} \rfloor]$ \\
$U_m \gets U^\prime[ \lfloor \frac{\Delta_b}{\Delta} \rfloor :  \lfloor \frac{\Delta_b}{\Delta} \rfloor + \lfloor \frac{\Delta_c}{\Delta} \rfloor]$ \\
$U_h \gets U^\prime[\lfloor \frac{\Delta_b}{\Delta} \rfloor + \lfloor \frac{\Delta_c}{\Delta} \rfloor:]$ \\
% Randomly sample from each partition based on respective budgets
$U_b \gets$ randomly sample $\Delta_b$ accounts from $U_l$ \\
$U_c \gets$ randomly sample $\Delta_c$ accounts from $U_m$ \\
$U_w \gets$ randomly sample $\Delta_w$ accounts from $U_h$ \\
\caption{Malicious Accounts Categorization}
\label{alg: malicious account categorization}
\end{algorithm}

Given the involvement of various malicious groups in misinformation campaigns \cite{shao2018spread, pacheco2021uncovering}, we model three distinct types of malicious accounts—bots, cyborgs, and crowd workers—characterized by varying influence levels and different budgets.
Using the budgets for the three malicious groups, denoted as $\Delta_b$, $\Delta_c$, and $\Delta_w$, we develop an adaptive categorization algorithm that generates the bot group $U_b$ with low influence, the cyborg group $U_c$ with medium influence, and the worker group $U_w$ with high influence. 
Specifically, we first sort all users in $U$ by the influence metric $\mathcal{I}$ in ascending order (line $1$ in Algorithm \ref{alg: malicious account categorization}). 
Next, we determine the sizes of the low, medium, and high influence groups according to the specified budgets and categorize all accounts into these groups (lines $2$-$5$ in Algorithm \ref{alg: malicious account categorization}). 
% We then categorize all accounts into the three groups based on their sorted influence levels (lines $4$-$6$ in Algorithm \ref{alg: malicious account categorization}).
Finally, we randomly sample accounts from each group to return $U_b$, $U_c$, and $U_w$ (lines $6$-$8$ in Algorithm \ref{alg: malicious account categorization}).
The controlled malicious accounts $U_m$ are thus defined as follows:
\begin{equation}
    U_m = U_b \cup U_c \cup U_w \text{.}
\end{equation}

Finally, to simulate the coordinated behavior among the different groups, we design three agents, each embodying a distinct level of influence: agent $\mathcal{N}_b$ for low-influence social bots, agent $\mathcal{N}_c$ for medium-influence cyborg accounts, and agent $\mathcal{N}_w$ for high-influence crowd workers.

\subsection{Target Subgraph Attack} \label{target subgraph attack}
Each attack on a target post, primarily focusing on fake news (though applicable to real news as well), is modeled as a collective effort within the associated user-post subgraph, where all agents work collaboratively to manipulate the classification outcome of a black-box GNN-based detector.

For a target fake news post $p \in P_t$, its associated user-post subgraph $G_\alpha$ includes the closely related posts $P_\alpha$, including both fake news posts $P_\alpha^f=\{p_1^f, p_2^f, \dots, p^f_{l_f}\}$ with $p=p_1^f$ and real news posts $P_\alpha^r=\{p_1^r, p_2^r, \dots, p_{l_r}\}$.
Here, $l_f$ and $l_r$ denote the number of fake and real news posts, respectively, within the post subset $P_\alpha \subset P$.

The SI2AF framework models the attack on the target news post $p$ as a cooperative multi-agent Markov decision process, characterized by the tuple $(\mathcal{N},\mathcal{S},\mathcal{A},\mathcal{P},\mathcal{R},\gamma)$, where $\mathcal{N}=\{\mathcal{N}_b, \mathcal{N}_c, \mathcal{N}_w\}$ is the set of agents, $\mathcal{S}$ denotes the state space observed by all agents, $\mathcal{A}$ is the joint action space, $\mathcal{P}$ represents the transition function, $\mathcal{R}$ refer to the reward function, and $\gamma$ is the discount factor.
At each timestep $t$, the agent $\mathcal{N}_b$, responsible for managing malicious accounts $U_b=\{u_1^b,u_2^b,\dots,u_{\Delta_b}^b\}$, observes the current environmental state $s_t \in \mathcal{S}$ and selects actions $\boldsymbol{a^b_t}=(a^b_1,a^b_2,\dots,a^b_{\Delta_b})$ according to its policy network $\pi_b$.
The policy network $\pi_b$ determines which post vertex each controlled account will interact with in the associated user-post subgraph $G_\alpha$, expressed as $\boldsymbol{a^b_t}=\pi_b(s_t,U_b,G_\alpha)$.
Similarly, agents $\mathcal{N}_c$ and $\mathcal{N}_w$ select their respective actions $\boldsymbol{a^c_t}$ and $\boldsymbol{a^w_t}$ using their own policy networks, following a decision-making process analogous to that of $\mathcal{N}_b$.

For each malicious account $u_i^b \in U_b$, we define the sampled probability $p^b_i$ of its selected action $a_i^b$ based on the cumulative entropy of all common parent nodes shared by $u_i^b$ and the target post $p \in P$ as follows:
\begin{equation} \label{equ: bot sample}
    p^b_i = \sum_{\{u_i^b, p\} \subset V_\alpha} H^{T^*_{up}}(G_{up};\alpha)\text{.}
\end{equation}
If the only common parent node between $u^b_i$ and $p$ is the root node, we set the sampled probability $p^b_i$ to a predefined random small value, $0.01$, to reflect a low likelihood of action.
Similarly, we define the sampled probabilities for the accounts controlled by agents $\mathcal{N}_c$ and $\mathcal{N}_w$ following the same approach.
Based on these probabilities, we perform a weighted sampling process on $\boldsymbol{a^b_t}$, $\boldsymbol{a^c_t}$, and $\boldsymbol{a^w_t}$, which leads to the single-agent actions $a^b_t$, $a^c_t$, and $a^w_t$ at timestep $t$.
Moreover, we centrally aggregate these actions, weighting them according to the sum of the network influences exerted by the malicious accounts controlled by each agent.
This aggregation yields the final action $a_t$ at timestep $t$, which specifies the attacked post $p_t \in P_\alpha$ and the selected malicious account $u_t \in U_m$.
The collective action $(p_t,u_t)$ modifies the structure of user-post graph $G_{up}$ by establishing a new sharing between $u_t$ and $p_t$, potentially affecting the classification outcome of target news $p$ by the GNN-based detector $f_{\theta^*}$.
Depending on the types of attacked post $p_t$, our subgraph attack encompasses three distinct attack strategies:

$\bullet$ \textbf{Direct Attack}: Directly interact with the target news, $p_t = p$, to affect its classification outcome by the GNN-based detector.

$\bullet$ \textbf{Indirect Attack}: Engage with real news within the associated subgraph, $p_t \in P_\alpha^r$, to indirectly affect the prediction of target $p$.

$\bullet$ \textbf{Feedback Attack}: Interact with other fake news in the associated subgraph, $p_t \in P_\alpha^f$ and $p_t \neq p$, with the aim to enrich the environmental feedback and address the challenge of reward sparsity in the decision process.

In its efforts to mount adversarial attacks against the GNN-based detectors $f_{\theta^*}$, the SI2AF framework considers the classification outcomes of the target post and other related fake news posts.
The predictions $\hat{y}_1,\hat{y}_2,\dots,\hat{y}_{l_f}$ for these posts act as reward signals, guiding the training and refinement of the policy networks of all agents in $\mathcal{N}$.
Specifically, $\hat{y}_1=f_{\theta^*}\left(X_p,E_{up}^\prime\right)_{p^f_1}$ represents the classification outcome for the target post $p = p_1^f$, while $\hat{y}_i=f_{\theta^*}\left(X_p,E_{up}^\prime\right)_{p^f_i}$ denotes the results for other fake posts $p^f_i$ where $i > 1$.
The reward $\mathcal{R}(s_t, a_t)$ is defined as follows: 
\begin{equation}
\mathcal{R}(s_t, a_t) = 
\begin{cases} 
1 & \text{if } f_{\theta^*}\left(X_p,E_{up}^\prime\right)_{p^f_1} \neq y_1 \text{,}\\
\sum_{i=2}^{l_f}{\frac{\mathbf{1}\left(f_{\theta^*}\left(X_p,E_{up}^\prime\right)_{p^f_i} \neq y_i\right)}{l_f - 1}} & \text{otherwise,}
\end{cases}
\end{equation}
where $E^\prime_{up}$ denotes the updated structural relationships perturbed by the action $a_t$.

For the agent $\mathcal{N}_b$ operating under policy $\pi_b$, we employ Q-learning to estimate its value function $\mathcal{Q}_b$ and minimize the optimization loss as follows:
\begin{equation} \label{equ: training loss}
    \mathcal{L}_{\mathcal{Q}_b} = \mathbb{E}_{(s_t,a_t^b)}{\left[\mathcal{R}(s_t,a^b_t) + \gamma\max_{a^b_{t+1}}\mathcal{Q}^-_b(s_{t+1},a^b_{t+1}) - \mathcal{Q}_b(s_t,a_t^b)\right]}\text{,}
\end{equation}
where $\mathcal{Q}^-_b$ denotes the target value network of agent $\mathcal{N}_b$, introduced to stabilize the training process by reducing oscillations in the learned Q-values.
The optimal value function $\mathcal{Q}^*_b(s_t,a^b_t)$ is expressed using the following Bellman Equation:
\begin{equation}
    \mathcal{Q}^*_b(s_t,a^b_t) = \mathcal{R}(s_t,a^b_t) + \gamma \max_{a^b_{t+1}}{\mathcal{Q}^*_b(s_{t+1},a^b_{t+1})}\text{.}
\end{equation}
This equation describes a greedy policy, where the agent $\mathcal{N}_b$ selects the action that maximizes the Q-value for the given state:
\begin{equation}
    \pi_b(\boldsymbol{a^b_t}|s_t;\mathcal{Q}^*_b) = \arg \max_{\boldsymbol{a^b_t}}{\mathcal{Q}^*_b(s_{t},\boldsymbol{a^b_{t+1}})} \text{.}
\end{equation}
The policy training for agents $\mathcal{N}_c$ and $\mathcal{N}_w$ follows the same Q-learning methodology as that of $\mathcal{N}_b$, with adaptations to their unique action spaces.

\begin{algorithm}[t]
\SetAlgoVlined
\KwIn{target post set $P_t$, trained policies $\pi_b^*, \pi_c^*, \pi_w^*$}
\KwOut{optimized detector $f_{\theta^*}$}
$E^\prime_{up} = E_{up}$ \\
\For{$p \in P_t$}{
$G_\alpha = (U, P_\alpha, E_\alpha^{up}) \gets$ derive the associated subgraph \\
$P^f_\alpha$ and $P^r_\alpha \gets$ extract the fake and real news in $G_\alpha$ \\
\While{$t < t_{max}$}{
$\boldsymbol{a^b_t} \gets \pi^*_b(s_t,U_b,G_\alpha)$ \\
$\boldsymbol{a^c_t} \gets \pi^*_c(s_t,U_c,G_\alpha)$ \\
$\boldsymbol{a^w_t} \gets \pi^*_w(s_t,U_w,G_\alpha)$ \\
$a_t^b$, $a_t^c$, and $a_t^w \gets$ individually sample $\boldsymbol{a^b_t}$, $\boldsymbol{a^c_t}$, and $\boldsymbol{a^w_t}$ via Equation \ref{equ: bot sample} \\ 
$a_t = (u_t,p_t) \gets$ sample single-agent actions $a^b_t$, $a^c_t$, and $a^w_t$ \\
$E^\prime_{up} = E^\prime_{up} \bigcup \{(u_t,p_t)\}$\\
}
}
$f_{\theta^*} \gets$ refine the graph-based detector on $E^\prime_{up}$ by minimizing the cross-entropy loss in Equation \ref{equ: ce loss}
\caption{Graph-based Detector Optimization}
\label{alg: graph-based optimization}
\end{algorithm}

\subsection{Detection Optimization}
By leveraging the trained SI2AF framework, we incorporate the generated manipulations within each subgraph to update the structural relationships between users and posts, thereby optimizing graph-based detectors to improve their robustness.
% This optimization process is summarized in Algorithm \ref{alg: graph-based optimization}.

For each target post $p \in P_t$, we extract the fake news $p^f_\alpha$ and real news $p^r_\alpha$ from the associated subgraph $G_\alpha$ (lines $3$ and $4$ in Algorithm \ref{alg: graph-based optimization}).
According to the trained policies $\pi^*_b$, $\pi^*_c$, and $\pi^*_w$, we select the multi-agent actions $\boldsymbol{a^b_t}$, $\boldsymbol{a^c_t}$, and $\boldsymbol{a^w_t}$, respectively, and employ weighted sampling to determine the collective action $a_t = (p_t, u_t)$ at timestep $t$ (lines $6$-$10$ in Algorithm \ref{alg: graph-based optimization}). 
This action is then used to update the structural relationships $E_{up}$ within the user-post graph $G_{up}$ (line $11$ in Algorithm \ref{alg: graph-based optimization}).
After all attacks targeting posts in $P_t$, we minimize the cross-entropy loss in Equation \ref{equ: ce loss} to refine the graph-based model $f_{\theta^*}$, thereby enhancing its detection robustness.

\begin{table*}[t]
\centering
\caption{The success rates of the SI2AF and other baselines on both fake and real news within the Politifact and Gossipcop datasets: ``average value $\pm$ standard deviation". \textbf{Bold}: the best performance in each graph, \underline{underline}: the second performance.}
\vspace{-0.3cm}
\resizebox{\textwidth}{!}{
\begin{tabular}{ccccccccccc}
\hline
\multirow{2}{*}{\textbf{Method}} & \multicolumn{5}{|c}{\textbf{Politifact Fake News}} & \multicolumn{5}{|c}{\textbf{Politifact Real News}} \\ \cline{2-11}
                        & \multicolumn{1}{|c}{\textbf{GAT}} & \multicolumn{1}{c}{\textbf{GCN}} & \multicolumn{1}{c}{\textbf{SAGE}} & \multicolumn{1}{c}{\textbf{Bi-GCN}} & \multicolumn{1}{c}{\textbf{GCAN}} & \multicolumn{1}{|c}{\textbf{GAT}} & \multicolumn{1}{c}{\textbf{GCN}} & \multicolumn{1}{c}{\textbf{SAGE}} & \multicolumn{1}{c}{\textbf{Bi-GCN}} & \multicolumn{1}{c}{\textbf{GCAN}} \\ \hline
Random & \multicolumn{1}{|c}{$0.14 \pm 0.01$} & \multicolumn{1}{c}{$0.09 \pm 0.03$} & \multicolumn{1}{c}{$0.13 \pm 0.01$} & \multicolumn{1}{c}{$0.10 \pm 0.02$} & \multicolumn{1}{c}{$0.09 \pm 0.01$} & \multicolumn{1}{|c}{$0.11 \pm 0.01$} & \multicolumn{1}{c}{$0.36 \pm 0.04$} & \multicolumn{1}{c}{$0.15 \pm 0.01$} & \multicolumn{1}{c}{$0.16 \pm 0.03$} & \multicolumn{1}{c}{$0.05 \pm 0.01$} \\
DICE & \multicolumn{1}{|c}{$0.26 \pm 0.02$} & \multicolumn{1}{c}{$0.16 \pm 0.01$} & \multicolumn{1}{c}{$0.24 \pm 0.03$} & \multicolumn{1}{c}{$0.15 \pm 0.01$} & \multicolumn{1}{c}{$0.17 \pm 0.03$} & \multicolumn{1}{|c}{$0.22 \pm 0.03$} & \multicolumn{1}{c}{$0.38 \pm 0.02$} & \multicolumn{1}{c}{$0.21 \pm 0.01$} & \multicolumn{1}{c}{$0.21 \pm 0.02$} & \multicolumn{1}{c}{$0.10 \pm 0.02$} \\ \hline
SGA & \multicolumn{1}{|c}{$0.32 \pm 0.04$} & \multicolumn{1}{c}{$0.24 \pm 0.01$} & \multicolumn{1}{c}{$0.35 \pm 0.03$} & \multicolumn{1}{c}{$\underline{0.21} \pm 0.01$} & \multicolumn{1}{c}{$\underline{0.27} \pm 0.01$} & \multicolumn{1}{|c}{$0.18 \pm 0.03$} & \multicolumn{1}{c}{$\underline{0.45} \pm 0.04$} & \multicolumn{1}{c}{$0.29 \pm 0.01$} & \multicolumn{1}{c}{$0.32 \pm 0.03$} & \multicolumn{1}{c}{$0.15 \pm 0.03$} \\
GAFSI & \multicolumn{1}{|c}{$\underline{0.35} \pm 0.03$} & \multicolumn{1}{c}{$\underline{0.28} \pm 0.02$} & \multicolumn{1}{c}{$\underline{0.36} \pm 0.02$} & \multicolumn{1}{c}{$0.19 \pm 0.02$} & \multicolumn{1}{c}{$0.26 \pm 0.04$} & \multicolumn{1}{|c}{$0.37 \pm 0.02$} & \multicolumn{1}{c}{$0.42 \pm 0.02$} & \multicolumn{1}{c}{$\underline{0.37} \pm 0.03$} & \multicolumn{1}{c}{$\underline{0.41} \pm 0.02$} & \multicolumn{1}{c}{$0.13 \pm 0.01$} \\ \hline
MARL & \multicolumn{1}{|c}{$0.30 \pm 0.01$} & \multicolumn{1}{c}{$0.21 \pm 0.01$} & \multicolumn{1}{c}{$0.35 \pm 0.03$} & \multicolumn{1}{c}{$0.12 \pm 0.02$} & \multicolumn{1}{c}{$0.18 \pm 0.02$} & \multicolumn{1}{|c}{$\underline{0.47} \pm 0.01$} & \multicolumn{1}{c}{$0.35 \pm 0.02$} & \multicolumn{1}{c}{$0.18 \pm 0.09$} & \multicolumn{1}{c}{$0.27 \pm 0.03$} & \multicolumn{1}{c}{$\underline{0.16} \pm 0.02$} \\ \hline
SI2AF(Ours) & \multicolumn{1}{|c}{$\textbf{0.41} \pm 0.03$} & \multicolumn{1}{c}{$\textbf{0.31} \pm 0.04$} & \multicolumn{1}{c}{$\textbf{0.41} \pm 0.01$} & \multicolumn{1}{c}{$\textbf{0.28} \pm 0.05$} & \multicolumn{1}{c}{$\textbf{0.34} \pm 0.01$} & \multicolumn{1}{|c}{$\textbf{0.69} \pm 0.04$} & \multicolumn{1}{c}{$\textbf{0.56} \pm 0.03$} & \multicolumn{1}{c}{$\textbf{0.49} \pm 0.02$} & \multicolumn{1}{c}{$\textbf{0.45} \pm 0.03$} & \multicolumn{1}{c}{$\textbf{0.19} \pm 0.01$} \\ \hline
Abs.($\%$) Avg.$\uparrow$ & \multicolumn{1}{|c}{$0.06(17.14\%)$} & \multicolumn{1}{c}{$0.03(10.71\%)$} & \multicolumn{1}{c}{$0.05(13.89\%)$} & \multicolumn{1}{c}{$0.07(33.33\%)$} & \multicolumn{1}{c}{$0.07(25.93\%)$} & \multicolumn{1}{|c}{$0.22(46.81\%)$} & \multicolumn{1}{c}{$0.11(24.44\%)$} & \multicolumn{1}{c}{$0.12(32.43\%)$} & \multicolumn{1}{c}{$0.04(9.76\%)$} & \multicolumn{1}{c}{$0.03(18.75\%)$} \\ \hline
\multirow{2}{*}{\textbf{Method}} & \multicolumn{5}{|c}{\textbf{Gossipcop Fake News}} & \multicolumn{5}{|c}{\textbf{Gossipcop Real News}} \\ \cline{2-11}
                        & \multicolumn{1}{|c}{\textbf{GAT}} & \multicolumn{1}{c}{\textbf{GCN}} & \multicolumn{1}{c}{\textbf{SAGE}} & \multicolumn{1}{c}{\textbf{Bi-GCN}} & \multicolumn{1}{c}{\textbf{GCAN}} & \multicolumn{1}{|c}{\textbf{GAT}} & \multicolumn{1}{c}{\textbf{GCN}} & \multicolumn{1}{c}{\textbf{SAGE}} & \multicolumn{1}{c}{\textbf{Bi-GCN}} & \multicolumn{1}{c}{\textbf{GCAN}} \\ \hline
Random & \multicolumn{1}{|c}{$0.18 \pm 0.01$} & \multicolumn{1}{c}{$0.25 \pm 0.03$} & \multicolumn{1}{c}{$0.15 \pm 0.01$} & \multicolumn{1}{c}{$0.17 \pm 0.02$} & \multicolumn{1}{c}{$0.33 \pm 0.02$} & \multicolumn{1}{|c}{$0.08 \pm 0.01$} & \multicolumn{1}{c}{$0.17 \pm 0.03$} & \multicolumn{1}{c}{$0.14 \pm 0.01$} & \multicolumn{1}{c}{$0.15 \pm 0.02$} & \multicolumn{1}{c}{$0.26 \pm 0.02$} \\
DICE & \multicolumn{1}{|c}{$0.13 \pm 0.02$} & \multicolumn{1}{c}{$0.10 \pm 0.01$} & \multicolumn{1}{c}{$0.16 \pm 0.02$} & \multicolumn{1}{c}{$0.24 \pm 0.03$} & \multicolumn{1}{c}{$0.29 \pm 0.02$} & \multicolumn{1}{|c}{$0.11 \pm 0.03$} & \multicolumn{1}{c}{$0.21 \pm 0.02$} & \multicolumn{1}{c}{$0.18 \pm 0.02$} & \multicolumn{1}{c}{$0.27 \pm 0.01$} & \multicolumn{1}{c}{$0.23 \pm 0.02$} \\ \hline
SGA & \multicolumn{1}{|c}{$0.42 \pm 0.02$} & \multicolumn{1}{c}{$0.72 \pm 0.03$} & \multicolumn{1}{c}{$0.19 \pm 0.03$} & \multicolumn{1}{c}{$0.33 \pm 0.01$} & \multicolumn{1}{c}{$0.53 \pm 0.02$} & \multicolumn{1}{|c}{$0.28 \pm 0.01$} & \multicolumn{1}{c}{$\underline{0.45} \pm 0.06$} & \multicolumn{1}{c}{$0.37 \pm 0.04$} & \multicolumn{1}{c}{$0.31 \pm 0.03$} & \multicolumn{1}{c}{$0.29 \pm 0.02$} \\
GAFSI & \multicolumn{1}{|c}{$0.21 \pm 0.01$} & \multicolumn{1}{c}{$0.67 \pm 0.04$} & \multicolumn{1}{c}{$\underline{0.20} \pm 0.04$} & \multicolumn{1}{c}{$\underline{0.49} \pm 0.03$} & \multicolumn{1}{c}{$0.61 \pm 0.03$} & \multicolumn{1}{|c}{$\underline{0.29} \pm 0.03$} & \multicolumn{1}{c}{$0.43 \pm 0.04$} & \multicolumn{1}{c}{$\underline{0.39} \pm 0.02$} & \multicolumn{1}{c}{$0.40 \pm 0.03$} & \multicolumn{1}{c}{$0.37 \pm 0.04$} \\ \hline
MARL & \multicolumn{1}{|c}{$\underline{0.80} \pm 0.04$} & \multicolumn{1}{c}{$\underline{0.78} \pm 0.02$} & \multicolumn{1}{c}{$0.13 \pm 0.01$} & \multicolumn{1}{c}{$0.28 \pm 0.05$} & \multicolumn{1}{c}{$\underline{0.78} \pm 0.09$} & \multicolumn{1}{|c}{$0.25 \pm 0.03$} & \multicolumn{1}{c}{$0.41 \pm 0.01$} & \multicolumn{1}{c}{$0.29 \pm 0.03$} & \multicolumn{1}{c}{$\underline{0.44} \pm 0.02$} & \multicolumn{1}{c}{$\underline{0.43} \pm 0.01$} \\ \hline
SI2AF(Ours) & \multicolumn{1}{|c}{$\textbf{0.90} \pm 0.02$} & \multicolumn{1}{c}{$\textbf{0.87} \pm 0.01$} & \multicolumn{1}{c}{$\textbf{0.21} \pm 0.01$} & \multicolumn{1}{c}{$\textbf{0.60} \pm 0.01$} & \multicolumn{1}{c}{$\textbf{0.88} \pm 0.02$} & \multicolumn{1}{|c}{$\textbf{0.32} \pm 0.01$} & \multicolumn{1}{c}{$\textbf{0.52} \pm 0.01$} & \multicolumn{1}{c}{$\textbf{0.42} \pm 0.01$} & \multicolumn{1}{c}{$\textbf{0.47} \pm 0.01$} & \multicolumn{1}{c}{$\textbf{0.46} \pm 0.01$} \\ \hline
Abs.($\%$) Avg.$\uparrow$ & \multicolumn{1}{|c}{$0.10(12.5\%)$} & \multicolumn{1}{c}{$0.09(11.54\%)$} & \multicolumn{1}{c}{$0.01(5.00\%)$} & \multicolumn{1}{c}{$0.11(22.45\%)$} & \multicolumn{1}{c}{$0.10(12.82\%)$} & \multicolumn{1}{|c}{$0.03(10.34\%)$} & \multicolumn{1}{c}{$0.07(15.56\%)$} & \multicolumn{1}{c}{$0.03(7.69\%)$} & \multicolumn{1}{c}{$0.03(6.82\%)$} & \multicolumn{1}{c}{$0.03(6.98\%)$} \\ \hline
\end{tabular}}
\label{table: success rate}
\vspace{-0.3cm}
\end{table*}

\begin{figure*}[t]
   \centering
    \includegraphics[width=1\textwidth]{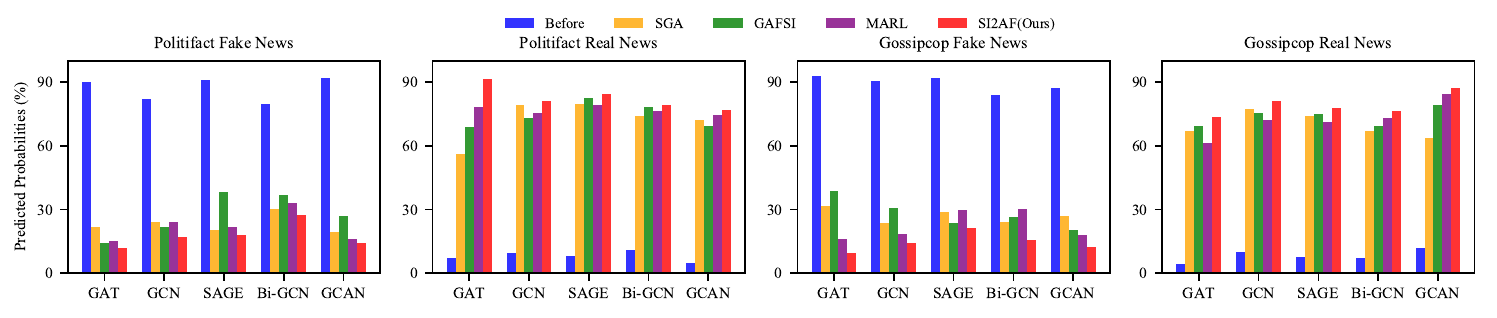}
    \vspace{-0.8cm}
    \caption{Average predictive probabilities of fake and real news before and after adversarial attacks.}
    \vspace{-0.4cm}
    \label{fig: predictive probability} 
\end{figure*}

\section{Experiments}\label{Experiments}
In this section, we conduct comprehensive comparative experiments on various real-world datasets to evaluate the performance of our proposed framework, SI2AF. 
To ensure a fair and robust assessment, the experimental results are reported as average values with standard deviations, calculated over five different random seeds.

\subsection{Experimental Settings}

\textbf{Datasets.}
For our analysis, we adopt two well-established real-world datasets, Politifact and Gossipcop, which originate from two fact-checking platforms and include social interactions from Twitter \cite{shu2020fakenewsnet}. 
These datasets contain metadata such as user interactions, post characteristics for fake and real news posts, and account information involved in these engagements.
In line with existing studies \cite{dou2021user}, we utilize Glove embeddings \cite{pennington2014glove} to encode both the semantic content of posts and the historical posts of users. 
Following the experimental settings \cite{wang2023attacking}, we adopt the same budget setting, randomly sampling $100$ bots, $50$ cyborgs, and $20$ crowd workers from the Politifact dataset, and $1000$ bots, $500$ cyborgs, and $100$ crowd workers from the Gossipcop dataset.

\noindent\textbf{Detection Models.}
In this study, we evaluate the effectiveness of SI2AF attack strategies against five different GNN-based detectors on the Politifact and Gossipcop datasets.
The GNN models employed as detectors include Graph Convolution Network (GCN) \cite{kipf2016semi}, Graph Attention Network (GAT) \cite{velivckovic2018graph}, Graph Sample and Aggregation Network (GraphSAGE) \cite{hamilton2017inductive}, Graph-aware Co-Attention Network (GCAN) \cite{lu2020gcan}, and Bi-Directional Graph Convolutional Network (Bi-GCN) \cite{bian2020rumor}.
Their detection performances after training are provided in Appendix \ref{app: detection performance}.

\noindent\textbf{Baselines.}
We compare SI2AF with several state-of-the-art baselines, including random-based methods (Random and DICE \cite{waniek2018hiding}), gradient-based methods (SGA \cite{li2021adversarial} and GASFI \cite{zhu2024general}), and the multi-agent cooperative method (MARL \cite{wang2023attacking}), using their publicly available open-source implementations.

\subsection{Evaluation}
To evaluate attack performance, we use the success rate as our primary metric, defined as the proportion of target posts successfully misclassified by the GNN-based detectors. 
We assess SI2AF and other baselines on their abilities to misclassify fake and real news in the Politifact and Gossipcop datasets, reporting the average success rate and standard deviation in Table \ref{table: success rate}.
Our experimental results show that SI2AF consistently outperforms all baselines, achieving maximum success rate improvements of up to $33.33\%$ for fake news and $46.81\%$ for real news across various attack scenarios.
% In the Politifact dataset, SI2AF increases the success rate of fake news attacks by [$0.06$, $0.03$, $0.05$, $0.07$, $0.07$] for the GAT, GCN, GraphSAGE, GCAN, and Bi-GCN models, respectively, and achieves [$0.22$, $0.11$, $0.12$, $0.04$, $0.03$] improvements for real news attacks.
% Similarly, SI2AF exhibits comparable performance advantages in the Gossipcop dataset, consistently outperforming the attack baselines and reinforcing its enhanced ability to evade detection.
For a deeper understanding of attack performance, we present the average predictive probabilities of target news posts in Figure \ref{fig: predictive probability}, illustrating the likelihood of these posts being classified as fake before and after attacks.
The results indicate that SI2AF induces more significant changes in predictive probabilities for fake and real posts, outperforming the best-performing baselines (SGA, GAFSI, and MARL) across different detection models. 
Specifically, SI2AF achieves an average reduction of $71.90\%$ in the predictive probabilities of fake posts and an increase of $72.90\%$ for real posts.

\begin{figure}[t]
   \centering
    \includegraphics[width=1\columnwidth]{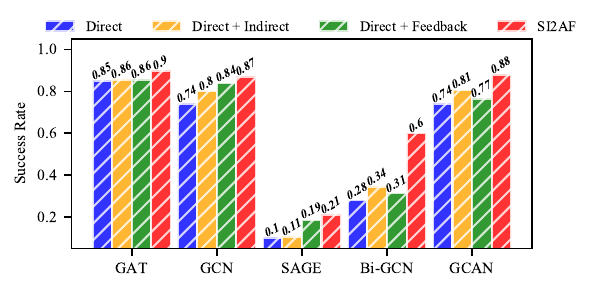}
    \vspace{-0.95cm}
    \caption{Success rates of different attack strategies on fake news in the Gossipcop dataset.}
    \vspace{-0.55cm}
    \label{fig: attack type} 
\end{figure}

As outlined in Section \ref{target subgraph attack}, the subgraph attack in SI2AF incorporates three distinct strategies, each targeting a specific category of news posts. 
In Figure \ref{fig: attack type}, we analyze the impact of selectively applying different attack strategies within the Gossipcop dataset and report the corresponding success rates. 
The combination of different attack strategies consistently yields higher success rates than using any single attack type alone, highlighting the strategic advantage and adaptability of the subgraph attack mechanism within SI2AF.
By adaptively deriving the hierarchical community structure for all accounts and posts, the SI2AF framework enables a more comprehensive and targeted subgraph attack on closely related news posts, significantly enhancing the attack's precision and effectiveness.

\begin{table}[t]
\centering
\caption{The detection performance of graph-based detectors against Gossipcop fake news before and after attacks.}
\vspace{-0.3cm}
\resizebox{\columnwidth}{!}{
\begin{tabular}{c|ccccc}
\hline
\multicolumn{1}{c}{\textbf{Method}} & \multicolumn{1}{|c}{\textbf{GAT}} & \multicolumn{1}{|c}{\textbf{GCN}} & \multicolumn{1}{|c}{\textbf{SAGE}} & \multicolumn{1}{|c}{\textbf{Bi-GCN}} & \multicolumn{1}{|c}{\textbf{GCAN}} \\ \hline 
Before & \multicolumn{1}{|c}{$93.1 \pm 0.5$} & \multicolumn{1}{|c}{$90.4 \pm 0.3$} & \multicolumn{1}{|c}{$91.8 \pm 0.3$} & \multicolumn{1}{|c}{$83.9 \pm 0.7$} & \multicolumn{1}{|c}{$87.3 \pm 0.4$} \\ \hline
SGA & \multicolumn{1}{|c}{$76.3 \pm 0.8$} & \multicolumn{1}{|c}{$64.1 \pm 0.5$} & \multicolumn{1}{|c}{$75.4 \pm 0.3$} & \multicolumn{1}{|c}{$62.8 \pm 0.3$} & \multicolumn{1}{|c}{$71.6 \pm 0.4$} \\
GAFSI & \multicolumn{1}{|c}{$72.9 \pm 1.0$} & \multicolumn{1}{|c}{$68.3 \pm 0.5$} & \multicolumn{1}{|c}{$65.3 \pm 0.6$} & \multicolumn{1}{|c}{$65.1 \pm 0.5$} & \multicolumn{1}{|c}{$60.7 \pm 0.4$} \\
MARL & \multicolumn{1}{|c}{$63.0 \pm 0.4$} & \multicolumn{1}{|c}{$68.4 \pm 0.7$} & \multicolumn{1}{|c}{$70.9 \pm 0.2$} & \multicolumn{1}{|c}{$61.4 \pm 0.6$} & \multicolumn{1}{|c}{$58.2 \pm 0.8$} \\
SI2AF & \multicolumn{1}{|c}{$56.4 \pm 0.6$} & \multicolumn{1}{|c}{$60.5 \pm 0.3$} & \multicolumn{1}{|c}{$61.8 \pm 0.2$} & \multicolumn{1}{|c}{$53.7 \pm 0.3$} & \multicolumn{1}{|c}{$56.2 \pm 0.5$} \\ \hline
\end{tabular}}
\label{table: optimization}
\vspace{-0.4cm}
\end{table}

In summary, our SI2AF framework leverages the structural information inherent in social networks to provide a more comprehensive and effective evaluation of robustness across various graph-based detectors.
Meanwhile, as illustrated in Algorithm \ref{alg: graph-based optimization}, we incorporate the generated manipulations from SI2AF to enrich the network structure and refine all five detectors on the updated network.
We summarize their detection performance on the Gossipcop fake news dataset, both before and after SI2AF attacks, alongside three best-performing baselines, in Table \ref{table: optimization}.
It is noted that, regardless of the graph-based detector used, the drop in predictive probability for each attack algorithm is significantly mitigated after optimization, with a reduction of at average $41.54\%$. 
The average prediction probability for fake news remains above $53.7\%$.
This is due to the comprehensive and strategic manipulations in SI2AF, which enable the detector to anticipate structural changes in the network caused by the attack algorithms, thereby significantly mitigating their impact on detection.

\begin{table}[t]
\centering
\caption{Efficiency comparison between SI2AF and MARL with different attack budgets.}
\vspace{-0.3cm}
\resizebox{\columnwidth}{!}{
\begin{tabular}{ccccc}
\hline
\multirow{3}{*}{\textbf{Methods}} & \multicolumn{4}{|c}{\textbf{Attack Budgets ($\Delta_b : \Delta_c : \Delta_w$)}} \\ \cline{2-5} 
& \multicolumn{2}{|c}{\textbf{100: 50: 20}} & \multicolumn{2}{|c}{\textbf{150: 75: 30}} \\ \cline{2-5} 
& \multicolumn{1}{|c}{\textbf{Training Time}} & \multicolumn{1}{c}{\textbf{Inference Time}} & \multicolumn{1}{|c}{\textbf{Training Time}} & \multicolumn{1}{c}{\textbf{Inference Time}} \\ \hline
MARL & \multicolumn{1}{|c}{$493.99$} & \multicolumn{1}{c}{$29.23$} & \multicolumn{1}{|c}{$548.36$} & \multicolumn{1}{c}{$52.64$} \\
SI2AF & \multicolumn{1}{|c}{$514.32$} & \multicolumn{1}{c}{$34.57$} & \multicolumn{1}{|c}{$560.41$} & \multicolumn{1}{c}{$59.16$}  \\ \hline
\multirow{3}{*}{\textbf{Methods}} & \multicolumn{4}{|c}{\textbf{Attack Budgets ($\Delta_b : \Delta_c : \Delta_w$)}} \\ \cline{2-5} 
& \multicolumn{2}{|c}{\textbf{200: 100: 40}} & \multicolumn{2}{|c}{\textbf{250: 125: 50}} \\ \cline{2-5} 
& \multicolumn{1}{|c}{\textbf{Training Time}} & \multicolumn{1}{c}{\textbf{Inference Time}} & \multicolumn{1}{|c}{\textbf{Training Time}} & \multicolumn{1}{c}{\textbf{Inference Time}} \\ \hline
MARL & \multicolumn{1}{|c}{$579.59$} & \multicolumn{1}{c}{$63.67$} & \multicolumn{1}{|c}{$586.70$} & \multicolumn{1}{c}{$71.04$} \\
SI2AF & \multicolumn{1}{|c}{$582.35$} & \multicolumn{1}{c}{$71.47$} & \multicolumn{1}{|c}{$594.28$} & \multicolumn{1}{c}{$75.33$}  \\ \hline
\end{tabular}}
\label{table: attack efficiency}
\vspace{-0.4cm}
\end{table}

To intuitively reflect the efficiency and scalability of our framework, we progressively increase the attack budget, that is, the number of malicious accounts controlled by the three agents, and record the time cost (ms) on single training or inference for both SI2AF and MARL in Table \ref{table: attack efficiency}.
Although incorporating the local structure around target news introduces additional computational overhead, the actual training and inference time of SI2AF remains comparable to that of MARL and remains stable as the budget increases, further demonstrating the practicality of our framework.

\begin{table}[t]
\centering
\caption{Attack performance of SI2AF and three best-performing baselines on posts with different engagements.}
\label{table: engagement}
\begin{tabular}{c|ccc}
\hline
\textbf{Post Degree} & \textbf{$[0, 10)$} & \textbf{$[10, 100)$} & \textbf{$[100, ~)$} \\ \hline
SGA & $0.39$ & $0.33$ & $0.21$ \\
GAFSI & $0.38$ & $0.35$ & $0.17$ \\
MARL & $0.38$ & $0.32$ & $0.24$ \\
SI2AF & $0.44$ & $0.40$ & $0.35$ \\ \hline
\end{tabular}
\vspace{-0.2cm}
\end{table}

To further explore the attack performance on posts with different engagements, we conduct additional analysis to evaluate the performance of our attack framework on target fake posts with varying levels of engagement. 
Specifically, we examine how the average success rate of fake posts varied based on their engagement levels, from newly posted, low-engagement content to high-engagement posts that had already spread significantly, using the SAGE detector on the Politifact dataset.
As shown in Table \ref{table: engagement}, compared to the baselines (SGA, GAFSI, and MARL), our method consistently achieves better attack performance across posts with different engagement levels. Notably, the advantage of our approach is more pronounced in high-degree posts, where attacking is more challenging due to their widespread engagement. 
This improvement is attributed to the richer and more diverse set of attack strategies we introduce, which are better suited to handle posts with varying degrees of influence.

\begin{figure*}[t]
   \centering
    \includegraphics[width=0.95\textwidth]{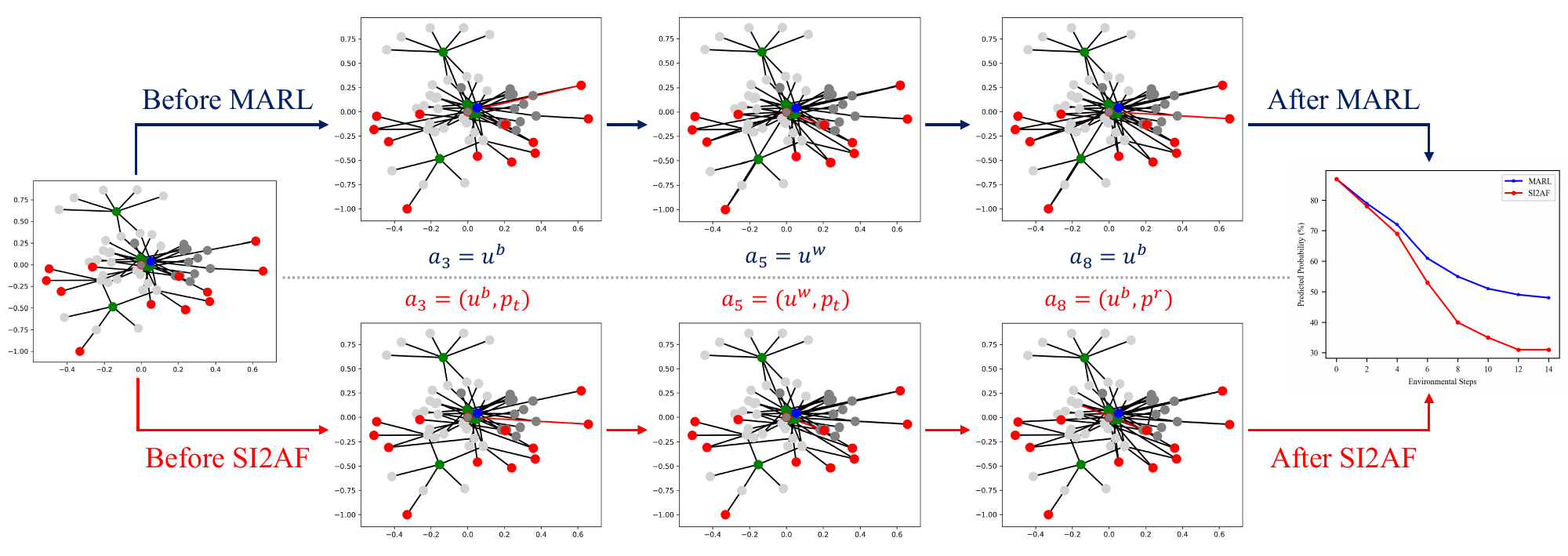}
    \vspace{-0.3cm}
    \caption{Attack visualization of SI2AF and MARL against GNN-based fake news detectors.}
    \vspace{-0.3cm}
    \label{fig: user case} 
\end{figure*}

\subsection{Case Study}
In this subsection, we focus on a specific fake post from the Gossipcop dataset and visualize the temporal changes in the associated user-post subgraph. 
We examine these changes induced by two multi-agent collaboration-based attack models: MARL and SI2AF. 
Finally, we compare the performance of these models based on their impact on the GNN-predicted probabilities.

To better visualize this process, we ensure that both attack models control the same set of malicious accounts, which are confined within their respective user-post communities. 
Figure \ref{fig: user case} illustrates that SI2AF initially selects malicious accounts in a manner similar to MARL, establishing direct connections with the target post. 
At this stage, SI2AF does not demonstrate a clear performance advantage over MARL in terms of modifying the GNN-predicted probability.
Subsequently, SI2AF expands its attack by connecting to other false and real posts, employing diverse strategies to influence the network. 
This leads to significant modifications in the target post’s GNN-based predicted probability, enhancing the overall effectiveness of the attack.

Further, we conduct a qualitative analysis of the SI2AF attack process, specifically focusing on how it influences the predicted probabilities of fake news in the context of the GNN-based model. We identify three primary mechanisms contributing to the increased predicted probabilities of fake news:

$\bullet$ Establishing new connections with influential malicious accounts, in accordance with our direct attack strategy.

$\bullet$ Amplifying the influence of malicious accounts already linked to the target post by connecting them to additional real news, consistent with our indirect attack strategy.

$\bullet$ Combining direct and indirect attacks to further increase the predicted probabilities of other fake news related to the target post, in line with our feedback strategy.

\begin{figure}[t]
   \centering
    \includegraphics[width=1\columnwidth]{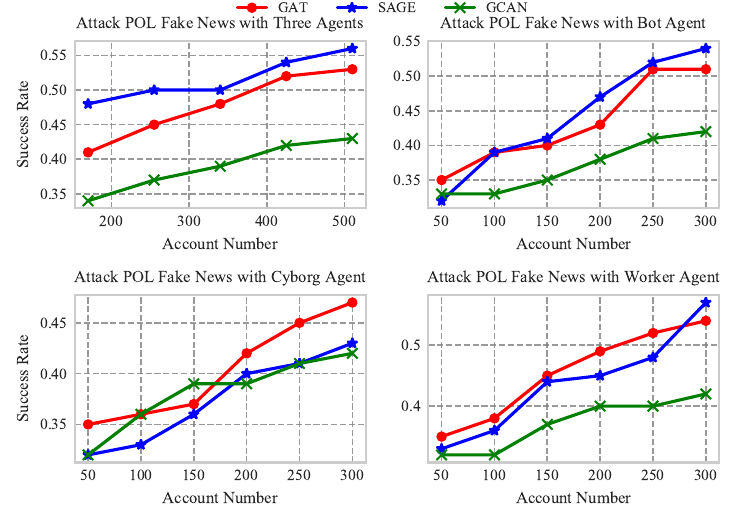}
    \vspace{-0.8cm}
    \caption{Attack performance of different agents on fake news detectors (GAT, SAGE, and GCAN) in the Politifact dataset.}
    \label{fig: abl_agent}
    \vspace{-0.5cm}
\end{figure}

\subsection{Ablation Studies}
In this subsection, we conduct an ablation study on the Politifact dataset for fake news detection to evaluate the impacts of various agents and their account quantities (ranging from $50$ to $300$) on the attack effectiveness within the SI2AF framework.
We focus on three of the most effective fake news detectors—GAT, SAGE, and GCAN—and gradually increase the number of accounts under each agent’s control to carry out the attack. 
Our results in Figure \ref{fig: abl_agent} suggest that increasing the number of accounts enhances attack performance for all agents up to a certain threshold, beyond which the improvement rate levels off.
Rather than improving performance, adding more accounts introduces more potential actions that may not contribute meaningfully to the attack strategy.
Notably, the worker agent consistently outperforms both the bot and cyborg agents across all three GNN detectors when controlling an equal number of accounts.

\begin{figure}[t]
   \centering
    \includegraphics[width=1\columnwidth]{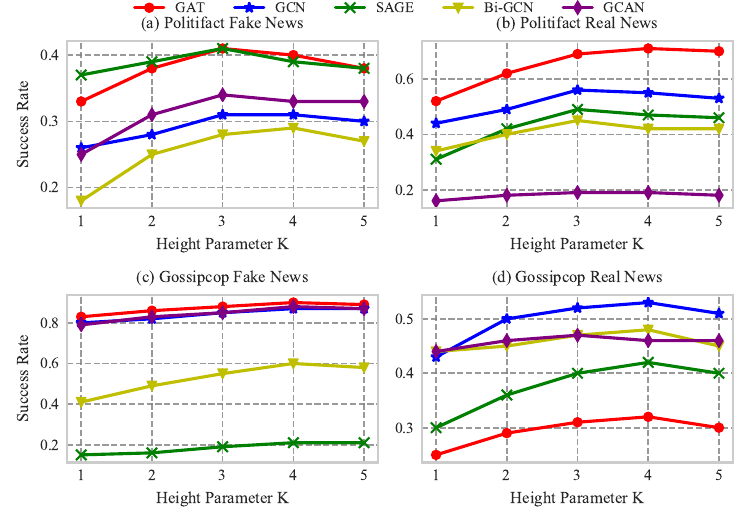}
    \vspace{-0.5cm}
    \caption{The success rate of SI2AF when adopting different height parameters $K$.}
    \vspace{-0.5cm}
    \label{fig: height parameter} 
\end{figure}

\subsection{Parameter Sensitivity}
In this subsection, we examine the attack performance of SI2AF with different height parameters, $K$, which controls the size of the subgraph considered in the targeted attack.
As shown in Figure \ref{fig: height parameter}, in both datasets, the attack success rate of SI2AF increases significantly as the value of parameter $K$ increases.
However, after reaching an optimal value, further increases in $K$ result in a slight performance decrease.
This decline occurs because a larger subgraph is more likely to include news posts unrelated to the target post, which dilutes the focus of the attack and reduces the effectiveness of SI2AF.
Additionally, the optimal value of $K$ is closely related to the social network scale. 
For instance, in the smaller network Politifact, SI2AF achieves its best performance when $K=3$, while in the larger-scale network Gossipcop, peak performance occurs when $K=4$.

\section{Conclusion}\label{Conclusion}
This paper proposes SI2AF, an adversarial attack framework that leverages network structural information to identify the hierarchical community structure among accounts and posts, thereby facilitating effective attacks against various GNN-based detectors and evaluating their robustness. 
We present an influence metric for categorizing malicious accounts, combined with three subgraph strategies utilizing multi-agent collaboration to maximize target news posts' evasion. 
Extensive experiments on two real-world datasets, Politifact and Gossipcop, demonstrate that SI2AF consistently enhances the attack effectiveness, outperforming state-of-the-art baselines, and significantly improves the robustness of graph-based detection.
Future research will focus on expanding the scope of graph-based detectors and enhancing their robustness through a more comprehensive exploration of subgraph attacks. 
Additionally, we plan to extend the categorization of malicious accounts and incorporate the dynamics of genuine accounts as key areas of investigation.

\begin{acks}
    This work has been supported by NSFC through grants 62322202, 62441612 and 62432006, Local Science and Technology Development Fund of Hebei Province Guided by the Central Government of China through grant 246Z0102G, the "Pioneer” and “Leading Goose” R\&D Program of Zhejiang" through grant 2025C02044, National Key Laboratory under grant 241-HF-D07-01, and Hebei Natural Science Foundation through grant F2024210008.
\end{acks}

%%
%% The next two lines define the bibliography style to be used, and
%% the bibliography file.
\bibliographystyle{ACM-Reference-Format}
\bibliography{main}

%%
%% If your work has an appendix, this is the place to put it.
\appendix
\newpage
\section{Framework Details}
\subsection{Primary Notations} \label{appendix: notations}
% \vspace{-0.3cm}
\begin{table}[h]
    \centering
    \caption{Notation Glossary}
    % \vspace{-0.3cm}
    \begin{tabular}{@{}l|l@{}}
    \toprule
        \textbf{Notation} & \textbf{Description} \\ \midrule
        $G=(V, E)$ & The undirected graph with vertex set and edge set. \\
        $v, e$ & The single vertex and undirected edge. \\
        % $d_v$ & The weighted sum of vertex degree. \\
        $H, T$ & The structural entropy and encoding tree. \\
        $T$ & The one- or high-layer encoding tree. \\
        $\lambda, \nu, \alpha$ & The root node, leaf nodes, and other nodes. \\
        % $l_\alpha$ & The number of $\alpha$ node's children. \\
        $g_\alpha, \mathcal{V}_\alpha$ &  The terms associated with the vertex subset $V_\alpha$. \\
        $K$ & The maximal height of the encoding trees. \\
        $\mathcal{I}, c$ & The influence metric and adjusting parameter. \\
        $x, q$ & The random variable and probability density function. \\
    \bottomrule
        $U, P$ & The sets of users and posts. \\
        $u, p$ & The single user account and single news post. \\
        $m, n$ & The respective numbers of users and posts. \\
        $X, Y$ & The feature matrix and label set. \\
        $h, y$ & The hidden representation and single label. \\
        % $w$ & The weight of an edge between user and post vertices. \\
        $f_\theta$ & The GNN prediction function parameterized by $\theta$. \\
        $\sigma, \Delta$ & The sigmoid function and attack budget. \\
    \bottomrule
        $\mathcal{N}$ & The set of multiple agents. \\
        $\mathcal{S}, \mathcal{A}$ & The state and action spaces. \\
        $\mathcal{P}, \mathcal{R}$ & The transition and reward functions. \\
        % $\gamma$ & The discount factor for expected reward. \\
        $s, a, r$ & The single state, action, and reward. \\
        $\pi, \mathcal{Q}$ & The policy network and value function. \\
    \bottomrule
    \end{tabular}
\end{table}

\subsection{Limitations}
In our study, we focus on attacking graph-based news detectors by manipulating the network structure and classify malicious accounts based on their structural characteristics and the influence they exert on the target post.
We draw on prior research \cite{wang2023attacking, shu2017fake} to define three types of malicious accounts—bots, cyborgs, and crowd workers.
These account types represent varying levels of influence within the network, with bots exerting low influence, cyborgs having medium influence, and crowd workers having high influence.
This classification is rooted in the idea that different account types interact with the network in distinct ways, and understanding their influence helps us design more targeted and effective attack strategies.
However, we acknowledge that malicious behaviors also include other actions such as spam, scams, and social engineering attacks.
These forms of maliciousness are important but are outside the scope of our current work, which focuses specifically on structural manipulation of the network.
In future research, we plan to expand our framework to include content-based features to capture a broader range of malicious accounts, including those engaged in spam or scam activities.
By incorporating these features, we aim to improve the generalizability of SI2AF and extend it to cover a wider variety of malicious behaviors.

On the other hand, we primarily simulate network dynamics around malicious accounts during the attack process to model certain dynamic aspects. However, we have not yet accounted for the dynamic behaviors of genuine users, which would further complicate the network structure during an attack.
We plan to extend our framework in future journal versions to handle dynamic network structures. This will include implementing mechanisms for real-time adjustments to attack strategies, allowing the framework to better align with the continuously evolving nature of social networks. By incorporating dynamic changes into our model, we aim to make the framework more applicable to real-world scenarios, where networks are constantly changing.

\subsection{Ethical Statement}
Our proposed adversarial attack framework targets the evaluation and enhancement of graph-based news detectors' robustness.
We explicitly disavow any unethical use of this framework and emphasize that our research is focused on strengthening detection systems for societal benefit.
 Our work aligns with the broader goal of promoting the integrity and security of AI-driven systems, particularly in the fight against misinformation and online manipulation.

\subsection{Time Complexity}
We outline SI2AF's attacking process against a fake news detector in Algorithm \ref{alg: SI2AF attacking}. 
The proposed SI2AF operates through several sequential stages: first, hierarchical structure identification for the user-post graph $G_{up}=(U, P, E_{up})$, which has a computational cost of $O(|E_{up}| + (|U| + |P|) \cdot \log^2 (|U| + |P|))$; second, multiple agent design with a time complexity of $O(|U| \cdot \log |U|)$; and finally, a subgraph attack for each target, where agents coordinate their actions in the joint action space of $|U_m| \cdot |P_\alpha|$, with $|P_\alpha|$ representing the number of posts in the associated subgraph $G_\alpha$.

\begin{algorithm}[h]
\SetAlgoVlined
\KwIn{maximal timesteps $t_{max}$, update interval $t_{up}$}
\KwOut{agent policies $\pi_b,\pi_c,\pi_w$}
% Hierarchical Structure Extraction
$G_{up} \gets$ construct the user-post graph \\
$T^*_{up} = \arg \min_{T_{up}}\{H^{T_{up}}(G_{up})\}$ \\
% Multiple Agent Definition
$U_m = U_b \cup U_c \cup U_w \gets$ categorize malicious accounts via Algorithm \ref{alg: malicious account categorization} \\
\For{$p \in P_t$} {
$E^\prime_{up} \gets E_{up}$ \\
$G_\alpha = (U, P_\alpha, E_\alpha^{up}) \gets$ derive the associated subgraph \\
$P^f_\alpha$ and $P^r_\alpha \gets$ extract the fake and real news in $G_\alpha$ \\
\While{$t < t_{max}$}{
$\boldsymbol{a^b_t} \gets \pi_b(s_t,U_b,G_\alpha)$ \\
$\boldsymbol{a^c_t} \gets \pi_c(s_t,U_c,G_\alpha)$ \\
$\boldsymbol{a^w_t} \gets \pi_w(s_t,U_w,G_\alpha)$ \\
$a_t^b$, $a_t^c$, and $a_t^w \gets$ individually sample $\boldsymbol{a^b_t}$, $\boldsymbol{a^c_t}$, and $\boldsymbol{a^w_t}$ via Equation \ref{equ: bot sample} \\ 
$a_t = (u_t,p_t) \gets$ sample single-agent actions $a^b_t$, $a^c_t$, and $a^w_t$ \\
$E^\prime_{up} = E^\prime_{up} \bigcup \{(u_t,p_t)\}$\\
\If{$t \mod t_{up} == 0$}{
$\pi_b$, $\pi_c$, and $\pi_w \gets$ update policies by minimizing loss in Equation \ref{equ: training loss}
}
}
}
\caption{SI2AF Attacking against Fake News Detector}
\label{alg: SI2AF attacking}
\end{algorithm}

\newpage
\section{Proof of Theorem \ref{theorem: 4.1}} \label{appendix: 4.1}
\begin{proof}
    Considering the transformation:
    \begin{equation}
        x^\prime = -\frac{x}{b} \cdot \log_{2} \left(\frac{c}{b} x\right) \text{,}
    \end{equation}
    which consists of a linear term, $-\frac{x}{b}$, and a logarithmic term, $\log_{2} (\frac{c}{b} x)$.
    For $x \in [1,\frac{b}{2}]$ and given the condition $0 < c \leq \frac{2}{e}$, the argument of the logarithm, $\frac{c}{b} x$, is strictly positive.
    Consequently, both $-\frac{x}{b}$ and $\log_{2} (\frac{c}{b} x)$ are continuous functions on $x \in [1,\frac{b}{2}]$.
    Hence, their product, $y$, is continuous over this interval.
    
    To examine the differentiability of $x^\prime$, we compute the derivative of variable $x^\prime$ with respect to $x$ using standard differentiation rules.
    Applying the product rule yields:
    \begin{equation} \label{equ: derivative}
        \frac{\mathrm{d}x^\prime}{\mathrm{d}x} = - \frac{1}{b} \cdot \left(\log_2 \frac{c}{b} + \log_2 x + \log_2 e\right) \text{.}
    \end{equation}
    Setting the derivative equal to zero to find critical points:
    \begin{equation}
        \frac{\mathrm{d}x^\prime}{\mathrm{d}x} = 0 \text{,}
    \end{equation}
    we obtain:
    \begin{equation}
        \log_2 \frac{c}{b} + \log_2 x + \log_2 e = 0 \text{,}
    \end{equation}
    which simplifies to:
    \begin{equation}
        x = \frac{b}{ec} \text{.}
    \end{equation}
    
    Given the constraints $0 < c \leq \frac{2}{e}$ and the fact that $1 \leq x \leq \frac{b}{2}$, we observe the following relationship:
    \begin{equation}
        x \leq \frac{b}{2} \leq \frac{b}{ec} \text{.}
    \end{equation}
    Therefore, within the interval $[1,\frac{b}{2}]$, the variable $x^\prime$ increases monotonically with the variables $x$, since the derivative $\frac{\mathrm{d}x^\prime}{\mathrm{d}x}$ remains positive, ensuring its monotonic behavior.

    Furthermore, the derivative $\frac{\mathrm{d}x^\prime}{\mathrm{d}x}$, as given by Equation \ref{equ: derivative}, is a monotonically decreasing function of $x$.
    We now calculate its minimum and maximum values over the interval $x \in [1,\frac{b}{2}]$ as follows:
    \begin{equation}
        \left(\frac{\mathrm{d}x^\prime}{\mathrm{d}x}\right)_{min} = {\left(\frac{\mathrm{d}x^\prime}{\mathrm{d}x}\right)}_{x=\frac{b}{2}} = \frac{1}{b} \cdot \log_2 \frac{2}{ec} \geq 0 \text{,}
    \end{equation}
    \begin{equation}
        \left(\frac{\mathrm{d}x^\prime}{\mathrm{d}x}\right)_{max} = {\left(\frac{\mathrm{d}x^\prime}{\mathrm{d}x}\right)}_{x=1} = \frac{1}{b} \cdot \log_2 \frac{b}{ec} \geq 0 \text{.}
    \end{equation}
    
    To derive the probability density function $q_1(x^\prime)$, we user the relationship between the probability density functions of $x$ and $x^\prime$:
    \begin{equation}
        q_1(x^\prime) = q_0(x) \left\lvert\frac{\mathrm{d}x}{\mathrm{d}x^\prime}\right\rvert \text{.}
    \end{equation}
    Since $\frac{\mathrm{d}x^\prime}{\mathrm{d}x} \geq 0$ over $x \in [1,\frac{b}{2}]$, we have:
    \begin{equation}
        0 \leq q_0(x)_{min} \cdot \left\lvert\frac{\mathrm{d}x}{\mathrm{d}x^\prime}\right\rvert_{min} \leq q_1(y) \leq q_0(x)_{max} \cdot \left\lvert\frac{\mathrm{d}x}{\mathrm{d}x^\prime}\right\rvert_{max} \text{.}
    \end{equation}
    Substituting the bounds for $q_0(x)$ and $\left\lvert\frac{\mathrm{d}x^\prime}{\mathrm{d}x}\right\rvert$, we obtain the inequality:
    \begin{equation}
        q_1(x^\prime) \leq \frac{b}{1 - \log_2 ec} \text{.}
    \end{equation}
    This completes the proof.
\end{proof}

\section{Evaluation Details}
\subsection{Datasets}
The statistics of benchmark datasets, Politifact and Gossipcop, are summarized in Table \ref{table: dataset statistics}.

\begin{table}[h]
\centering
\caption{Dataset statistics.}
\label{table: dataset statistics}
\resizebox{\columnwidth}{!}{
\begin{tabular}{c|ccccc}
\hline
\textbf{Datasets} & \textbf{Nodes} & \textbf{Users} & \textbf{Posts} & \textbf{Edges} & \textbf{Targets} \\ \hline
Politifact & 276,858 & 276,277 & 581 & 1,074,890 & 62 \\ \hline
Gossipcop & 575,993 & 565,660 & 10,333 & 3,084,931 & 1547 \\ \hline
\end{tabular}}
\end{table}

\subsection{Detectors.} \label{app: detection performance}
We train all GNN-based detectors, GAT, GCN, SAGE, Bi-GCN, and GCAN, to optimize their balanced and accurate detection capabilities for fake and real news posts, using metrics accuracy and F1-score, as detailed in Table \ref{table: detectors}.

\begin{table}[h]
\centering
\caption{GNN-based detection performance.}
\resizebox{\columnwidth}{!}{
\begin{tabular}{ccccc}
\hline
\multirow{2}{*}{\textbf{Detection Model}} & \multicolumn{2}{|c}{\textbf{Politifact Dataset}}    & \multicolumn{2}{|c}{\textbf{Gossipcop Dataset}}     \\ \cline{2-5} 
                        & \multicolumn{1}{|c}{\textbf{Accuracy}} & \textbf{F1} & \multicolumn{1}{|c}{\textbf{Accuracy}} & \textbf{F1} \\ \hline
GCN & \multicolumn{1}{|c}{$0.8157$} & \multicolumn{1}{c}{$0.8024$} & \multicolumn{1}{|c}{$0.9383$} & \multicolumn{1}{c}{$0.9348$} \\
GAT & \multicolumn{1}{|c}{$0.8354$} & \multicolumn{1}{c}{$0.8340$} & \multicolumn{1}{|c}{$0.9316$} & \multicolumn{1}{c}{$0.9266$} \\
GraphSAGE & \multicolumn{1}{|c}{$0.8108$} & \multicolumn{1}{c}{$0.8102$} & \multicolumn{1}{|c}{$0.9252$} & \multicolumn{1}{c}{$0.9206$} \\
GCAN & \multicolumn{1}{|c}{$0.8475$} & \multicolumn{1}{c}{$0.8465$} & \multicolumn{1}{|c}{$0.9142$} & \multicolumn{1}{c}{$0.9081$} \\
Bi-GCN & \multicolumn{1}{|c}{$0.8084$} & \multicolumn{1}{c}{$0.8052$} & \multicolumn{1}{|c}{$0.8916$} & \multicolumn{1}{c}{$0.8840$} \\ \hline
\end{tabular}}
\label{table: detectors}
\end{table}

\section{Framework Scalability}
To validate the scalability of our framework for large social networks, we have developed a strategy to partition the network into local subgraphs based on vertex connectivity.
This approach facilitates parallel processing of multiple local subgraphs, which reduces overall time complexity and improves scalability.
We perform detailed analyses of training and inference times (ms) using a larger-scale Weibo dataset \cite{ma2016detecting}, as shown in the table \ref{table: time analysis}.
The results demonstrate that, despite the additional overhead from hierarchical community identification and subgraph targeting, the time costs introduced by our framework remain comparable to those of the original attack algorithms.
Even for large social networks, the parallelized processing ensures that computational overhead stays within acceptable limits.

\begin{table}[h]
\centering
\caption{Time analysis of our SI2AF and MARL baseline in the Weibo dataset.}
\label{table: time analysis}
\begin{tabular}{c|cc}
\hline
\textbf{Method} & \textbf{Training Time} & \textbf{Inference Time} \\ \hline
MARL & $537.29$ & $32.65$ \\
SI2AF & $542.35$ & $36.17$ \\ \hline
\end{tabular}
\end{table}

\begin{table*}[t]
\centering
\caption{Performance Comparison among single-agent variants of MARL and SI2AF within the Politifact and Gossipcop datasets.}
\vspace{-0.2cm}
\resizebox{\textwidth}{!}{
\begin{tabular}{cccccccccccc}
\hline
\multicolumn{1}{c}{\multirow{2}{*}{\textbf{Agent}}} & \multicolumn{1}{|c}{\multirow{2}{*}{\textbf{Method}}} & \multicolumn{5}{|c}{\textbf{Politifact Fake News}} & \multicolumn{5}{|c}{\textbf{Politifact Real News}} \\ \cline{3-12}
\multicolumn{1}{c}{} & \multicolumn{1}{|c}{} & \multicolumn{1}{|c}{\textbf{GAT}} & \multicolumn{1}{c}{\textbf{GCN}} & \multicolumn{1}{c}{\textbf{SAGE}} & \multicolumn{1}{c}{\textbf{Bi-GCN}} & \multicolumn{1}{c}{\textbf{GCAN}} & \multicolumn{1}{|c}{\textbf{GAT}} & \multicolumn{1}{c}{\textbf{GCN}} & \multicolumn{1}{c}{\textbf{SAGE}} & \multicolumn{1}{c}{\textbf{Bi-GCN}} & \multicolumn{1}{c}{\textbf{GCAN}} \\ \hline
\multicolumn{1}{c}{\multirow{2}{*}{Bot}} & \multicolumn{1}{|c}{MARL} & \multicolumn{1}{|c}{$0.34 \pm 0.01$} & \multicolumn{1}{c}{$\textbf{0.22} \pm 0.01$} & \multicolumn{1}{c}{$0.36 \pm 0.03$} & \multicolumn{1}{c}{$\textbf{0.22} \pm 0.04$} & \multicolumn{1}{c}{$0.26 \pm 0.05$} & \multicolumn{1}{|c}{$0.16 \pm 0.01$} & \multicolumn{1}{c}{$0.50 \pm 0.01$} & \multicolumn{1}{c}{$\textbf{0.32} \pm 0.08$} & \multicolumn{1}{c}{$0.36 \pm 0.04$} & \multicolumn{1}{c}{$0.07 \pm 0.01$} \\
\multicolumn{1}{c}{} & \multicolumn{1}{|c}{SI2AF} & \multicolumn{1}{|c}{$\textbf{0.37} \pm 0.03$} & \multicolumn{1}{c}{$0.20 \pm 0.01$} & \multicolumn{1}{c}{$\textbf{0.39} \pm 0.04$}  & \multicolumn{1}{c}{$\textbf{0.22} \pm 0.03$} & \multicolumn{1}{c}{$\textbf{0.31} \pm 0.01$} & \multicolumn{1}{|c}{$\textbf{0.35} \pm 0.07$} & \multicolumn{1}{c}{$\textbf{0.51} \pm 0.05$} & \multicolumn{1}{c}{$0.26 \pm 0.02$} & \multicolumn{1}{c}{$\textbf{0.41} \pm 0.01$} & \multicolumn{1}{c}{$\textbf{0.15} \pm 0.03$} \\ \hline
\multicolumn{1}{c}{\multirow{2}{*}{Cyborg}} & \multicolumn{1}{|c}{MARL} & \multicolumn{1}{|c}{$\textbf{0.33} \pm 0.02$} & \multicolumn{1}{c}{$\textbf{0.22} \pm 0.01$} & \multicolumn{1}{c}{$0.33 \pm 0.01$} & \multicolumn{1}{c}{$\textbf{0.12} \pm 0.01$} & \multicolumn{1}{c}{$0.16 \pm 0.02$} & \multicolumn{1}{|c}{$0.16 \pm 0.01$} & \multicolumn{1}{c}{$\textbf{0.50} \pm 0.02$} & \multicolumn{1}{c}{$\textbf{0.39} \pm 0.02$} & \multicolumn{1}{c}{$0.27 \pm 0.03$} & \multicolumn{1}{c}{$0.10 \pm 0.01$} \\
\multicolumn{1}{c}{} & \multicolumn{1}{|c}{SI2AF} & \multicolumn{1}{|c}{$0.20 \pm 0.03$} & \multicolumn{1}{c}{$0.19 \pm 0.01$} & \multicolumn{1}{c}{$\textbf{0.38} \pm 0.03$} & \multicolumn{1}{c}{$0.10 \pm 0.01$} & \multicolumn{1}{c}{$\textbf{0.23} \pm 0.02$} & \multicolumn{1}{|c}{$\textbf{0.47} \pm 0.15$} & \multicolumn{1}{c}{$\textbf{0.50} \pm 0.04$} & \multicolumn{1}{c}{$0.29 \pm 0.02$} & \multicolumn{1}{c}{$\textbf{0.33} \pm 0.02$} & \multicolumn{1}{c}{$\textbf{0.17} \pm 0.02$} \\ \hline
\multicolumn{1}{c}{\multirow{2}{*}{Worker}} & \multicolumn{1}{|c}{MARL}& \multicolumn{1}{|c}{$\textbf{0.34} \pm 0.02$} & \multicolumn{1}{c}{$\textbf{0.21} \pm 0.01$} & \multicolumn{1}{c}{$0.33 \pm 0.01$} & \multicolumn{1}{c}{$0.12 \pm 0.02$} & \multicolumn{1}{c}{$0.16 \pm 0.01$} & \multicolumn{1}{|c}{$0.16 \pm 0.01$} & \multicolumn{1}{c}{$0.48 \pm 0.02$} & \multicolumn{1}{c}{$0.32 \pm 0.08$} & \multicolumn{1}{c}{$0.27 \pm 0.01$} & \multicolumn{1}{c}{$0.10 \pm 0.01$} \\
\multicolumn{1}{c}{} & \multicolumn{1}{|c}{SI2AF} & \multicolumn{1}{|c}{$0.31 \pm 0.05$} & \multicolumn{1}{c}{$0.20 \pm 0.06$} & \multicolumn{1}{c}{$\textbf{0.37} \pm 0.02$} & \multicolumn{1}{c}{$\textbf{0.21} \pm 0.01$} & \multicolumn{1}{c}{$\textbf{0.30} \pm 0.01$} & \multicolumn{1}{|c}{$\textbf{0.29} \pm 0.03$} & \multicolumn{1}{c}{$\textbf{0.49} \pm 0.04$} & \multicolumn{1}{c}{$\textbf{0.35} \pm 0.08$} & \multicolumn{1}{c}{$\textbf{0.43} \pm 0.02$} & \multicolumn{1}{c}{$\textbf{0.16} \pm 0.03$} \\ \hline
\multicolumn{1}{c}{\multirow{2}{*}{\textbf{Agent}}} & \multicolumn{1}{|c}{\multirow{2}{*}{\textbf{Method}}} & \multicolumn{5}{|c}{\textbf{Gossipcop Fake News}} & \multicolumn{5}{|c}{\textbf{Gossipcop Real News}} \\ \cline{3-12}
\multicolumn{1}{c}{} & \multicolumn{1}{|c}{} & \multicolumn{1}{|c}{\textbf{GAT}} & \multicolumn{1}{c}{\textbf{GCN}} & \multicolumn{1}{c}{\textbf{SAGE}} & \multicolumn{1}{c}{\textbf{Bi-GCN}} & \multicolumn{1}{c}{\textbf{GCAN}} & \multicolumn{1}{|c}{\textbf{GAT}} & \multicolumn{1}{c}{\textbf{GCN}} & \multicolumn{1}{c}{\textbf{SAGE}} & \multicolumn{1}{c}{\textbf{Bi-GCN}} & \multicolumn{1}{c}{\textbf{GCAN}} \\ \hline
\multicolumn{1}{c}{\multirow{2}{*}{Bot}} & \multicolumn{1}{|c}{MARL} & \multicolumn{1}{|c}{$\textbf{0.82} \pm 0.02$} & \multicolumn{1}{c}{$0.29 \pm 0.03$} & \multicolumn{1}{c}{$0.14 \pm 0.01$} & \multicolumn{1}{c}{$0.22 \pm 0.02$} & \multicolumn{1}{c}{$\textbf{0.73} \pm 0.04$} & \multicolumn{1}{|c}{$0.18 \pm 0.05$} & \multicolumn{1}{c}{$0.40 \pm 0.10$} & \multicolumn{1}{c}{$0.32 \pm 0.01$} & \multicolumn{1}{c}{$\textbf{0.43} \pm 0.03$} & \multicolumn{1}{c}{$\textbf{0.4} \pm 0.01$} \\
\multicolumn{1}{c}{} & \multicolumn{1}{|c}{SI2AF} & \multicolumn{1}{|c}{$0.73 \pm 0.02$} & \multicolumn{1}{c}{$\textbf{0.36} \pm 0.02$} & \multicolumn{1}{c}{$\textbf{0.19} \pm 0.02$} & \multicolumn{1}{c}{$\textbf{0.5} \pm 0.01$} & \multicolumn{1}{c}{$0.49 \pm 0.02$} & \multicolumn{1}{|c}{$\textbf{0.30} \pm 0.01$} & \multicolumn{1}{c}{$\textbf{0.47} \pm 0.01$} & \multicolumn{1}{c}{$\textbf{0.34} \pm 0.01$} & \multicolumn{1}{c}{$0.41 \pm 0.01$} & \multicolumn{1}{c}{$0.39 \pm 0.03$} \\ \hline
\multicolumn{1}{c}{\multirow{2}{*}{Cyborg}} & \multicolumn{1}{|c}{MARL} & \multicolumn{1}{|c}{$\textbf{0.81} \pm 0.01$} & \multicolumn{1}{c}{$0.68 \pm 0.02$} & \multicolumn{1}{c}{$\textbf{0.13} \pm 0.03$} & \multicolumn{1}{c}{$0.08 \pm 0.01$} & \multicolumn{1}{c}{$0.69 \pm 0.06$} & \multicolumn{1}{|c}{$0.23 \pm 0.03$} & \multicolumn{1}{c}{$\textbf{0.45} \pm 0.01$} & \multicolumn{1}{c}{$0.32 \pm 0.01$} & \multicolumn{1}{c}{$\textbf{0.42} \pm 0.05 $} & \multicolumn{1}{c}{$\textbf{0.41} \pm 0.02$} \\
\multicolumn{1}{c}{} & \multicolumn{1}{|c}{SI2AF} & \multicolumn{1}{|c}{$0.71 \pm 0.06$} & \multicolumn{1}{c}{$\textbf{0.81} \pm 0.03$} & \multicolumn{1}{c}{$0.10 \pm 0.01$} & \multicolumn{1}{c}{$\textbf{0.53} \pm 0.03$} & \multicolumn{1}{c}{$\textbf{0.70} \pm 0.02$} & \multicolumn{1}{|c}{$\textbf{0.27} \pm 0.04$} & \multicolumn{1}{c}{$0.37 \pm 0.02$} & \multicolumn{1}{c}{$\textbf{0.37} \pm 0.03$} & \multicolumn{1}{c}{$0.35 \pm 0.02$} & \multicolumn{1}{c}{$0.21 \pm 0.09$} \\ \hline
\multicolumn{1}{c}{\multirow{2}{*}{Worker}} & \multicolumn{1}{|c}{MARL} & \multicolumn{1}{|c}{$\textbf{0.82} \pm 0.02$} & \multicolumn{1}{c}{$0.48 \pm 0.03$} & \multicolumn{1}{c}{$0.13 \pm 0.02$} & \multicolumn{1}{c}{$0.22 \pm 0.01$} & \multicolumn{1}{c}{$0.71 \pm 0.03$} & \multicolumn{1}{|c}{$\textbf{0.27} \pm 0.03$} & \multicolumn{1}{c}{$0.45 \pm 0.05$} & \multicolumn{1}{c}{$0.31 \pm 0.01$} & \multicolumn{1}{c}{$0.12 \pm 0.02$} & \multicolumn{1}{c}{$0.30 \pm 0.02$} \\
\multicolumn{1}{c}{} & \multicolumn{1}{|c}{SI2AF} & \multicolumn{1}{|c}{$0.76 \pm 0.03$} & \multicolumn{1}{c}{$\textbf{0.54} \pm 0.02$} & \multicolumn{1}{c}{$\textbf{0.16} \pm 0.01$} & \multicolumn{1}{c}{$\textbf{0.51} \pm 0.03$} & \multicolumn{1}{c}{$\textbf{0.73} \pm 0.03$} & \multicolumn{1}{|c}{$0.14 \pm 0.02$} & \multicolumn{1}{c}{$\textbf{0.48} \pm 0.04$} & \multicolumn{1}{c}{$\textbf{0.32} \pm 0.09$} & \multicolumn{1}{c}{$\textbf{0.39} \pm 0.03$} & \multicolumn{1}{c}{$\textbf{0.37} \pm 0.02$} \\ \hline
\end{tabular}}
\label{table: single agent}
\vspace{-0.15cm}
\end{table*}

\section{Single-agent Variant Comparison}
We compare the single-agent variants of our SI2AF framework with the multi-agent baseline, MARL \cite{wang2023attacking}.
Each variant controls a distinct group of malicious accounts (bots, cyborgs, or crowd workers) to execute attacks aimed at misclassifying fake and real news.
Table \ref{table: single agent} summarizes the average attack success rates and standard deviations across all five GNN-based detectors.

In the Bot and Worker variants of SI2AF, we observed that SI2AF outperforms MARL in nearly $80\%$ of attack scenarios, demonstrating a clear performance advantage. 
However, in the Cyborg variant, the performance difference between SI2AF and MARL is less pronounced. This variation can be attributed to two key factors: the classification of malicious accounts and the subgraph attack strategy employed by SI2AF.

The MARL baseline relies on local degree features for account classification, assuming that accounts with higher degrees have more influence. 
However, due to the heavy-tailed degree distribution in social networks, this approach results in a relatively fixed distribution of Worker accounts, which tend to have higher degrees. 
Consequently, Worker accounts are less adaptable and struggle to effectively attack target accounts with fewer structural connections, leading to weaker attack performance. 
In contrast, SI2AF uses global structural information for account classification, resulting in a more balanced distribution of Worker accounts and, consequently, better attack performance.

In the Bot variant, SI2AF’s subgraph attack strategy, including indirect and feedback attacks, is particularly effective. 
These strategies work well for Bot accounts, which have larger budgets and stronger internal collaboration, providing more flexibility in influencing the network. 
This is why the performance difference between SI2AF and MARL is more pronounced in the Bot variant.

To clarify, although the performance of SI2AF varies across agent types (Bots, Cyborgs, and Workers), the agents do not operate on entirely disjoint sets of predictions. 
Instead, each agent performs optimally based on its specific characteristics (such as account influence and budget). 
However, this does not imply that each agent works with completely separate sets of predictions; rather, each agent can influence predictions in different ways, depending on its unique influence within the network.

\end{document}